# Analyzing the Gelation Transition through Fractional Viscoelasticity and the Mittag-Leffler-Prabhakar Function

Yogesh M. Joshi[1, 2]

[1] Department of Chemical Engineering, Indian Institute of Technology Kanpur, Kanpur, Uttar Pradesh 208016, India.

[2] Centre for Nanosciences, Indian Institute of Technology Kanpur, Kanpur, Uttar Pradesh 208016, India.

* Email: joshi@iitk.ac.in

**Abstract**

The gelation transition, a process that transforms a flowable liquid into an elastic solid, occurs in a variety of systems, ranging from colloidal to polymeric. During the gelation transition, a system passes through a critical gel state characterized by scale-free power-law viscoelasticity. Interestingly, the fractional calculus provides a natural mathematical language for such power-law viscoelasticity. In this work, we develop physically constrained fractional viscoelastic models, as well as those based on the three-parameter Mittag-Leffler-Prabhakar function, for both the pre-gel and post-gel regimes, ensuring consistency with the conventional scaling relations in each regime. While the fractional pre-gel model is observed to be valid only for a restricted subset of parameter values, the Prabhakar function-based model for the relaxation modulus, which translates into a Havriliak–Negami function-based complex modulus, rigorously removes this limitation. We enforce continuity of the dynamic moduli and their derivatives across the critical gel point, which universally imposes a symmetry in the relaxation dynamics on either side of the critical gel state. Such enforcement further validates the hyperscaling relation connecting the critical exponents, making it a theoretical necessity rather than an empirical coincidence. We validate the proposed models against time- and frequency-domain experimental data. A model-agnostic, frequency-independent rheological fingerprint of the critical gel state, uniquely determined by two critical exponents, is also identified.

## Introduction

Soft matter spans a diverse class of materials whose mechanical response lies between that of an ideal elastic solid and a viscous liquid [1]. Among these, materials undergoing gelation transition, either sol-gel or gel-sol, occupy a particularly important position as it transforms viscoelastic liquid into a viscoelastic solid or vice versa. In addition, gels serve as model systems for probing the fundamental physics of network formation and have a vast range of applications [2, 3]. Broadly, gels can be classified into two principal categories: chemical gels and physical gels, each distinguished by the nature of the bond that characterizes the network architecture. Chemical gels typically consist of molecular units interconnected by permanent covalent crosslinks. Physical gels, on the other hand, may comprise polymers as well as colloidal particles. In physical gels the network is formed through associations such as hydrogen bonding, hydrophobic interactions, depletion forces, van der Waals attractions, electrostatic interactions, etc. [1]. Despite this fundamental disparity in their microstructural origins, chemical and physical gels undergoing the gelation transition exhibit striking similarities in their macroscopic rheological signatures. Such equivalence is attributed to universal features of network percolation rather than the specific chemistry or colloidal physics of any given system [4-6]. The remarkable feature of the gelation transition is the presence of a unique state, called the critical gel state, for which various linear viscoelastic response functions show a power dependence on their respective independent variable, time or frequency. Significant work has been reported in the literature that models how linear viscoelasticity evolves during the gelation transition [7-11]. Interestingly, a fractional viscoelastic element, such as a spring-pot, intrinsically leads to power-law viscoelasticity, making it a natural representation of the critical gel state [12]. Much work has also been done to model the gelation transition through a fractional viscoelasticity approach. Having said that, the gelation transition is a much more profound phenomenon. As the critical gel state is approached from the sol state, viscosity diverges to infinity, while as the material transitions beyond the critical gel state into the post-gel state, the equilibrium (zero-frequency) modulus continues to increase [7-11]. In addition, how viscosity diverges in the pre-gel (sol) state, the emergence of the equilibrium modulus in the post-gel state, and the power-law index associated with the critical gel state are all related through hyperscaling relations [5]. Therefore, there is a need to integrate hyperscaling relations with fractional viscoelasticity. We address this important topic in the present manuscript by proposing a representative model based on fractional viscoelasticity that adheres to the hyperscalimg relations.

The sol-gel and gel-sol transitions are of considerable practical consequence across a wide spectrum of applications [2, 3]. The foregoing discussion highlights that, notwithstanding systems as well as corresponding application domains, the universal rheological characteristics of the sol–gel transition remain largely agnostic to the specific pathways by which this transformation occurs. Nonetheless, the way the system gets driven across the transition plays a decisive role in how these signatures manifest and are experimentally probed. Establishing this connection between universal response and the underlying route to gelation provides a necessary context for interpreting the evolution of viscoelasticity across the transition. The process of sol-gel transition, interestingly, is extremely versatile as it can be induced through a variety of external stimuli or internal physicochemical processes. The nature of the variable that induces the gelation transition profoundly influences the reversibility of the same. In chemically crosslinking systems, it is well known that the covalent bonds accumulate irreversibly between reactive functional groups, and the corresponding sol-gel transition is permanent [4]. In physically associating systems, the gelation transition is driven by changes in time [13-18], temperature [19-28], radiation [29, 30], pH [31], or the concentration of a specific species [32-34], thereby enabling considerable external tunability. However, the physically associating systems may or may not be reversible. More specifically, many systems in which a change in temperature drives the sol-gel transition may undergo a gel-sol transition upon reversing the direction of temperature change [5, 22, 23, 26]. This thermoreversibility makes such systems uniquely suited for studying the sol-gel transition repeatedly or for probing both directions of the transition within a single material system. Some systems that show spontaneous (time-dependent) sol-gel transition may undergo gel-sol transition upon application of shear [16, 17] or freezing [35]. However, repeated application of shear in each system may not always pass through the unique critical gel state [36]. Such shear-induced gel-sol transition, followed by a spontaneous sol-gel transition, whether passing through the critical gel state or not, falls within the realm of thixotropy and has been discussed elsewhere [37].

The change in crosslinking density as the gelation progresses is an essential indicator of the extent of network formation. The conventional framework, originally developed for chemically crosslinking polymeric systems, introduces a parameter called degree of crosslinking, $p$, defined as the fraction of cross-linkable functional groups that have participated in the formation of a crosslinked network, so that $p \in [0,1]$ [4]. The degree of crosslinking at which the nascent network first percolates across the sample, also known as the critical gel point, corresponds to a value $p_c$. The distance from this critical state is then quantified by the dimensionless parameter $\varepsilon = |p - p_c|$, which vanishes precisely at the critical gel point and grows monotonically as the state departs into either the pre-gel ($p < p_c$, sol state) or the post-gel ($p > p_c$, gel state) regime. However, in practice, it is important to relate $\varepsilon$ to the experimentally accessible control parameter. Let us consider that parameter $X$ controls the gelation, and its value at the critical point is given by $X_c$. For spontaneous gelation, time is the controlling parameter, and hence we can take $X = t$, while for thermally driven gelation, we can consider $X = T$. Accordingly, $\varepsilon$ is defined as $\varepsilon \sim |X - X_c|/X_c$. [5, 14, 16, 38, 39]

Crucially, while the physical identity of the exact control variable $X$ changes the intrinsic nature of the systems entirely, the rheological response in the vicinity of the critical gel state obeys remarkably similar scaling forms. This affirms the notion of universality in the viscoelastic fingerprint of the gelation transition that will be discussed below.

The definition of the reduced distance $\varepsilon$ and the observed universality of the viscoelastic response is with respect to the reference intermediate critical gel state. At this state, by definition, the material exhibits scale-free viscoelasticity, wherein all linear response functions follow identical power-law scaling in the time and frequency domain, without any characteristic relaxation time [40]. In Table 1 we summarize this behavior in Eqs. (1) to (5), that describe expressions for relaxation modulus, dynamic moduli, and $\tan\delta$ at the critical gel state. As expressed, the relaxation modulus decays as a power-law with coefficient $n$. The dynamic moduli, on the other hand, scale with frequency with the identical coefficient, resulting in a frequency-independent loss tangent. Such a power-law dependence of moduli, as well as the frequency invariance of $\tan\delta$, provides a stringent experimental criterion for identifying the critical gel, which reflects the self-similar, fractal nature of the underlying nascent network. On either side of the critical gel state are the sol (pre-gel) and gel (post-gel) states. The departure from the critical gel state in either state introduces a characteristic timescale into the system. As the material approaches the critical gel state from the sol side, although it remains a viscoelastic liquid, its relaxation spectrum broadens, leading to divergence in viscosity and the longest relaxation time, as expected, respectively, by Eqs. (6) and (8). Conversely, in the post-gel state, the emergence of a percolated network gives rise to a finite equilibrium modulus that increases as the material departs further from the critical state, as shown by Eq. (7). The corresponding longest relaxation time associated with the finite clusters is given by Eq. (9).

**Table 1.** Summary of the linear viscoelastic relations across the gelation transition

| **Critical gel state** $[p = p_c, \varepsilon = \lvert p - p_c \rvert = 0]$ | | Eq. No. |
|---|---|---|
| Relaxation Modulus | $G(t) = St^{-n}, 0 < n < 1$ | (1) |
| Complex Modulus | $G^*(\omega) = S\Gamma(1-n)(i\omega)^n$ | (2) |
| Storage Modulus | $G'(\omega) = S\Gamma(1-n)\cos\left(\frac{n\pi}{2}\right)\omega^n$ | (3) |
| Loss Modulus | $G''(\omega) = S\Gamma(1-n)\sin\left(\frac{n\pi}{2}\right)\omega^n$ | (4) |
| Loss tangent | $\tan\delta = \tan\left(\frac{n\pi}{2}\right),$ | (5) |

| **Property** | **Pre-gel state** ($p < p_c$) $\varepsilon = \lvert p - p_c \rvert$ | Eq. No. | **Post-gel state** ($p > p_c$) $\varepsilon = \lvert p - p_c \rvert$ | Eq. No. |
|---|---|---|---|---|
| Characteristic quantity | Zero-shear viscosity: $\eta_0 = \eta_S \varepsilon^{-s}$ | (6) | Equilibrium modulus: $G_e = G_0 \varepsilon^{z}$ | (7) |
| Longest finite relaxation time | $\tau_{\max,S} = \tau_S\, \varepsilon^{-1/\kappa_S}$ | (8) | $\tau_{\max,G} = \tau_G\, \varepsilon^{-1/\kappa_G}$ | (9) |
| Scaling relation | $\kappa_S = \dfrac{1-n}{s}$ | (10) | $\kappa_G = \dfrac{n}{z}$ | (11) |
| Relaxation modulus $G(t,\varepsilon)$ | $G(t,\varepsilon) = St^{-n}\Phi\left(\dfrac{t}{\tau_{max,S}}\right)$ <br> $\Phi(x) = \begin{cases} 1, & x \to 0 \\ 0, & x \to \infty \end{cases}$ | (12) | $G(t,\varepsilon) = St^{-n}\Psi\left(\dfrac{t}{\tau_{max,G}}\right) + G_e(\varepsilon)$ <br> $\Psi(x) = \begin{cases} 1, & x \to 0 \\ 0, & x \to \infty \end{cases}$ | (13) |
| | $\left(\dfrac{\partial \ln G(t,\varepsilon)}{\partial \varepsilon}\right)_{\varepsilon \to 0} = -B_S t^{\kappa_S}$ | (14) | $\left(\dfrac{\partial \ln(G(t,\varepsilon) - G_e(\varepsilon))}{\partial \varepsilon}\right)_{\varepsilon \to 0} = B_G t^{\kappa_G}$ | (15) |
| Complex modulus $G^*(\omega,\varepsilon)$ | $\left(\dfrac{\partial \ln(G^*(\omega,\varepsilon))}{\partial \varepsilon}\right)_{\varepsilon \to 0} \sim \omega^{-\kappa_S}$ | (16) | $\left(\dfrac{\partial \ln(G^*(\omega,\varepsilon) - G_e(\varepsilon))}{\partial \varepsilon}\right)_{\varepsilon \to 0} \sim \omega^{-\kappa_G}$ | (17) |
| Proportionality between $G'$ and $G''$ | $\left(\dfrac{\partial \ln G'}{\partial \varepsilon}\right)_{\varepsilon \to 0} = C_{S\to G}\left(\dfrac{\partial \ln G''}{\partial \varepsilon}\right)_{\varepsilon \to 0}$ | (18) | $\left(\dfrac{\partial \ln(G' - G_e)}{\partial \varepsilon}\right)_{\varepsilon \to 0} = C_{G\to S}\left(\dfrac{\partial \ln G''}{\partial \varepsilon}\right)_{\varepsilon \to 0}$ | (19) |
| Linear viscoelastic continuity as $\varepsilon \to 0$ | $G^*(\omega,p)\rvert_{p\to p_c^-} = G^*(\omega,p)\rvert_{p\to p_c^+} = S\Gamma(1-n)(i\omega)^n$ | | | (20) |
| | $\left(\dfrac{\partial G^*(\omega,p)}{\partial p}\right)_{p\to p_c^-} = \left(\dfrac{\partial G^*(\omega,p)}{\partial p}\right)_{p\to p_c^+}$ | | | (21) |
| Hyper scaling relation | $\kappa_S = \kappa_G = \kappa$ | | | (22) |
| | $n = \dfrac{z}{z+s}$ | | | (23) |
| $\left(\dfrac{\partial G''}{\partial G'}\right)_{\varepsilon \to 0}$ | $C = C_{G\to S} = C_{S\to G} = \left(\dfrac{\partial \ln G'(\omega,\varepsilon)}{\partial \ln G''(\omega,\varepsilon)}\right)_{\varepsilon \to 0} = \cot\left(\dfrac{(n-\kappa)\pi}{2}\right)\tan\left(\dfrac{n\pi}{2}\right)$ | | | (24) |

The manner in which viscosity and longest relaxation time in the pre-gel regime, while equilibrium modulus and longest finite relaxation time in the post-gel regime evolve, is governed by a set of scaling relations, which interrelate the critical exponents across the transition as described by Eqs. (10) and (11) in Table 1. The corresponding admissible forms of the relaxation modulus on either side of the transition are described by Eqs. (12) and (13) in Table 1. An additional important experimental signature is the logarithmic rate of change of the complex and dynamic moduli with respect to the control parameter $p$ is described by Eqs. (16) to (19) in Table 1, which exhibit a characteristic power-law dependence on frequency, illustrating the underlying relaxation dynamics.

Joshi [40] identified the critical aspect connecting the pre-gel and post-gel regimes: the continuity of dynamic moduli as the system passes through the critical gel state, given by Eq. (21). This continuity condition imposes stringent constraints that unify the pre- and post-gel descriptions, and results in symmetry in relaxation dynamics ($\kappa_S = \kappa_G = \kappa$) on the either side of the critical gel state as described by Eq. (22) of Table 1. This condition, according to Joshi [40] is no longer an assumption but a strict requirement to validate continuity of dynamic moduli and their derivatives as the material undergoes gelation transition. Very interestingly, this continuity constraint, therefore, enforces the hyperscaling relation: $n = z/(z+s)$ (Eq. (23)) as a theoretical necessity rather than an empirical observation.[40] Interestingly, the validity of the symmetry condition of the relaxation dynamics and the hyperscaling relation also leads to an expression for the intriguing parameter $C$, given by $C = \left(\frac{\partial \ln G'(\omega,\varepsilon)}{\partial \ln G''(\omega,\varepsilon)}\right)_{\varepsilon\to 0}$ whose expression is given by Eq. (24) of Table 1. Joshi compared experimentally determined value of $C$ with that obtained from Eq. (24) with the knowledge of experimentally obtained $n$ and $\kappa$ for spontaneous gelation of cellulose nanocrystals.[17] Remarkably the expression of $C = \left(\frac{\partial \ln G'(\omega,\varepsilon)}{\partial \ln G''(\omega,\varepsilon)}\right)_{\varepsilon\to 0}$ is model agnostic. In appendix A, we present a derivation of $\left(\frac{\partial \ln G'(\omega,\varepsilon)}{\partial \ln G''(\omega,\varepsilon)}\right)_{\varepsilon\to 0}$ leading to Eq. (24) using Kramers-Kronig relations. The derivation is carried out separately for the pre-gel and post-gel states, starting from Eqs. (16) and (18) for the pre-gel states and Eqs. (17) and (19) for the post-gel states. Appendix A further shows that, independent of consideration of any constitutive model, continuity at the critical gelation transition point given by $C_{S\to G} = C_{G\to S}$ intrinsically leads to the hyperscaling relation: $n = z/(z+s)$. The exponents $n$ and $\kappa$ are obtained from experiment. In principle, they are related to the structure and dynamics of the clusters. The critical exponent $n$ is related to the fractal dimension $d_f$ of the critical cluster. For screened interactions in $d = 3$, Muthukumar [41] obtained

$$n = \frac{3\left(5-2d_f\right)}{2\left(5-d_f\right)} \qquad \Leftrightarrow \qquad d_f = \frac{5}{2}\,\frac{3-2n}{3-n}. \tag{25}$$

The exponent $\kappa$ describes how the longest relaxation time grows with the distance from the critical gel state and is given by Eq. (8). Other exponents, such as z and $s$, are related to $n$ and $\kappa$ through hyperscaling relations.

For a system undergoing sol-gel transition, Zaccone et al. [42] addressed a complementary question regarding how clusters produce a scale-free spectrum from the bottom up for attractive colloids. They consider a cluster of $k$ particles having size $R \sim k^{1/d_f}$, where $d_f$ is its mass fractal dimension. Such cluster relaxes by diffusing over its own size, resulting in its relaxation time $(\tau(k))$ scale as $\tau(k) \sim R^3 \sim k^{3/d_f}$. They propose a kinetic equation for aggregation and bond breakage, leading to the number $c_k$ of clusters of each size. Since each cluster size relaxes at its own time, the relaxation spectrum is essentially the cluster-size distribution rewritten in terms of $\tau$. When bonds can break (thermal breakage of dangling particles), their analysis leads, in the limit of long aggregation time, to $c_k \sim k^{-3/2}e^{-k/k_c}$ and therefore $H(\tau) \sim \tau^{-d_f/2}\exp\left[-(\tau/\tau_c)^{d_f/3}\right]$. At the gel point, since $\tau_c \to \infty$, they retrieve the Winter–Chambon power-law with $n = d_f/2$. However, when bonds cannot break, the corresponding spectrum is purely stretched-exponential, devoid of any power-law regime ($n = 0$). Since the relaxation modulus and the relaxation spectrum are mathematically interconvertible, the expression of Zaccone et al. [42] and Joshi [40] (as described in Table 1) leads to a power-law for $\tau \ll \tau_{\max}$ followed by a cut-off. In the case of Zaccone et al. [42], the shape of the cut-off follows from the aggregation kinetics of Brownian colloids. In the case of Joshi [40], the shape of the cut-off is that of the chosen model, which simultaneously satisfies the conditions at the gel point, including scaling relations and the continuity condition.

Interestingly, fractional calculus, a branch of mathematical analysis, when used to analyze viscoelasticity, intrinsically yields power-law rheology associated with the critical gel state [12, 13, 43]. This natural correspondence between the power-law rheology and fractional viscoelasticity was identified decades ago by Scott Blair [44]. The elementary mechanical element of fractional viscoelasticity that generates a pure power-law response is represented as a springpot (sometimes also referred to as the Scott Blair element). The constitutive stress ($\sigma(t)$) – strain ($\gamma(t)$) relationship for a springpot is given by [45, 46]:

$$\sigma(t) = \mathbb{V}\,\frac{d^\alpha \gamma}{dt^\alpha}, 0 \le \alpha \le 1, \tag{26}$$

where $\mathbb{V}$ is a quasi-property with units of $\mathrm{Pa\,s}^\alpha$ and $d^\alpha/dt^\alpha$ is the Caputo fractional derivative of order $\alpha$, which is bounded by 0 and 1. In the conventional viscoelastic approach, the state given by Eq. (26) has been represented by a hierarchy consisting of springs along one of the struts and dashpots on the rungs of the ladder, as proposed by Schiessel et al. [45]. The use of fractional calculus with a non-local, inherently scale-invariant power-law memory kernel, termed a springpot, essentially bypasses the intricate mathematical analysis of the aforementioned infinite assembly. By changing the value of $\alpha$, the springpot interpolates continuously between a Hookean spring ($\alpha =$

0) and a Newtonian dashpot ($\alpha = 1$). Eq. (26) can be easily solved to obtain the corresponding relaxation modulus and complex modulus of the springpot and is respectively given by:

$$G(t) = \frac{\mathbb{V}}{\Gamma(1-\alpha)}\, t^{-\alpha}, \tag{27}$$

$$G^*(\omega) = \mathbb{V}(i\omega)^{\alpha}. \tag{28}$$

Eqs. (27) and (28) are structurally identical to the relaxation modulus and complex modulus associated with the critical gel state, respectively given by: $G(t) = St^{-n}$ and $G^*(\omega) = S\Gamma(1-n)(i\omega)^n$ originally identified by Winter and Chambon [4, 47] with the correspondence $\alpha = n$ and $\mathbb{V} = S\Gamma(1-n)$. This equivalence establishes the springpot as the minimal fractional viscoelastic model of the critical gel state. Consequently, the springpot serves as the foundational element upon which more elaborate fractional viscoelastic models that show gelation transition have been constructed. It should be noted that, from a molecular perspective, the springpot is not merely a mathematical abstraction but a convenient form for the macroscopic manifestation of fractal relaxation dynamics. In classical polymer physics, the relaxation of a single polymer chain is often described by a discrete spectrum of modes. Particularly, the relaxation dynamics associated with the Rouse or Zimm models naturally lead to a power-law form at intermediate timescales (e.g., $G(t) \sim t^{-1/2}$ for Rouse dynamics) [1, 48]. During gelation, chemical crosslinks or physical associations cause monomer groups or particles to assemble into polydisperse, highly branched clusters. At the critical gel state, these clusters become self-similar and percolate the space associated with the macroscopic dimensions of the sample. The fractional derivative intrinsically captures this scale-free, fractal hierarchy of relaxation times. Accordingly, the fractional order $\alpha = n$ leading to $G(t) \sim t^{-n}$ encodes the hierarchical relaxation spectrum of the critical percolated network. There are several excellent monographs that give introduction to fundamentals on fractional viscoelasticity and its application to real material behavior [12, 43, 45, 46, 49-51]. Lately, fractional viscoelasticity is being used extensively to model the rheological behavior of soft materials [49, 52-61]. In this manuscript, we shall not go through the basics of fractional viscoelasticity, and readers are encouraged to refer to the above-mentioned monographs.

With this background, a hierarchy of fractional viscoelastic models has been developed by combining springpots in series as well as in parallel, analogous to the classical spring-dashpot networks in conventional linear viscoelasticity. Such springpot assemblies are then considered with or without springs and/or dashpots, depending upon the nature of the viscoelastic material being modeled. Typically, a dashpot in series with a springpot assembly naturally introduces a characteristic timescale and therefore produces a response that transitions from power-law solid behavior at short times to liquid-like terminal flow at long times, making it a natural candidate for describing pre-gel states. On the other hand, the parallel assembly of springpot and spring in the fractional Kelvin-Voigt model introduces a finite equilibrium modulus at long times and thereby captures the essential characteristic feature of the post-gel states. More general ladder network models and hierarchical arrangements of springpots along with springs and dashpots have also been proposed in the literature, which lead to richer relaxation spectra that can approximate arbitrary power-law viscoelastic behavior over extended frequency ranges [45]. A comprehensive account of such combinations of fractional and conventional viscoelastic elements and their application to a range of soft matter systems, including gels, has also been provided in the literature [12, 43, 45, 46, 49-51].

In the literature, a variety of fractional viscoelastic models have been employed to model the sol-gel transition. In particular, springpot-based models have been used to characterize the rheological state at or near the critical gel point, leveraging the direct equivalence between the springpot exponent and the critical relaxation exponent $n$. Keshavarz et al. [62] used a fractional Kelvin-Voigt framework to study nonlinear viscoelasticity of polymer gels near the gel point. Their model captured the linear viscoelastic envelope, thereby enabling systematic quantification of the departure from linear response. In an important contribution, Aime et al. [13] studied the power-law viscoelasticity of colloidal gels of Ludox® silica nanoparticles that undergo spontaneous, time-dependent gelation. They showed that fractional viscoelastic models provide a pertinent representation of the relaxation dynamics, which, through the fractional exponent, encode the self-similar structure of the percolating network. Keshavarz et al. [63] further demonstrated the potential of the fractional viscoelastic framework in the context of time-connectivity superposition for weak colloidal gels. They employed a springpot-based model to collapse dynamic moduli data obtained at different stages of network formation, construct master curves, and establish a direct connection between the fractional exponent and the evolving network microstructure. Morlet-Decarnin et al. [16, 17] studied critical-like gelation dynamics in cellulose nanocrystal suspensions using fractional viscoelastic models, further validating the utility of the springpot as a diagnostic tool for analyzing gelation in time-driven colloidal systems. These studies and the growing literature that use fractional viscoelasticity to characterize the approach to the critical gel state and the power-law relaxation dynamics collectively establish it as a valuable tool for rheological analysis.

Despite significant work on applying fractional viscoelasticity to the gelation transition, a careful examination of the existing literature reveals an important, hitherto unaddressed limitation. The most fractional viscoelastic models developed to date have been applied predominantly at the critical gel state, where the single springpot description is exact, or have been proposed for the individual states on either side of the transition without considering continuity of rheological evolution as the system undergoes the gelation transition through the critical point. Furthermore, to the best of our knowledge, no existing fractional viscoelastic framework has been formulated that enforces the explicit requirement that the model parameters and their evolution with $\varepsilon$ must satisfy the scaling and hyperscaling relations that are central to the gelation transition. More specifically, as proposed by Joshi [40], the continuity of the derivatives of the dynamic moduli with respect to the degree of crosslinking $p$ as the material passes through the critical gel point, as shown by Eq. (21), is a physical necessity. This continuity requirement enforces the symmetry condition $\kappa_S = \kappa_G = \kappa$ and the hyperscaling relation $n = z/(z + s)$. The fractional viscoelastic models proposed in the literature, as of now, have not been constructed to validate this continuity requirement given by Eq. (21), and consequently, none of them demonstrate adherence to hyperscaling relations as their intrinsic feature. This gap is the primary motivation for the present work, in which we propose fractional viscoelastic models that capture the evolution of rheological properties as a system undergoes the gelation transition. The models have been explicitly constructed to satisfy the continuity condition at the critical gel point, as well as the scaling and hyperscaling relations established in critical gelation theory.

## Model Development

In this section, we develop fractional viscoelastic model for relaxation modulus and the complex modulus of a material undergoing a gelation transition. We independently develop models for the pre-gel and post-gel states. We enforce two types of constraints on the models. The first type of constraints is those that are independently applicable in the pre-gel and post-gel states. These constraints, in addition to intrinsic thermodynamic compulsions, lead to the validation of the scaling relations in their respective domains. The second type of constraints concerns the continuity of dynamic moduli and their derivatives as the material passes through the critical gel state. As identified by Joshi [40], this is an essential constraint, as it enforces the relationships between the parameters of the model on either side of the critical gel state, strongly emphasizing that the evolution of linear viscoelasticity on either side of the critical gel state is not independent of the other.

The first set of constraints is intrinsic to the physics of viscoelastic materials. These constraints must be satisfied independently on each side of the transition. Specifically, any admissible model must satisfy:

$$\begin{aligned} G(t,\varepsilon) &= 0 \qquad t < 0, \\ G(t,\varepsilon) &\geq 0 \qquad t > 0. \end{aligned} \tag{29}$$

This constraint originates in causality. Next, in the limit of vanishing distance from the critical gel state, the relaxation modulus must recover the Winter-Chambon power-law expression given by [4],

$$\lim_{\varepsilon \to 0} G(t, \varepsilon) = S t^{-n}, \quad 0 < n < 1. \tag{30}$$

Then, for a fixed $\varepsilon$, the relaxation modulus must be a monotonically non-increasing function of time,

$$\left(\frac{\partial G(t, \varepsilon)}{\partial t}\right)_{\varepsilon} \leq 0, \tag{31}$$

which is a thermodynamic requirement. Furthermore, the Fourier transform of the relaxation modulus must yield non-negative storage and loss moduli for all frequencies,

$$G'(\omega, \varepsilon) \geq 0, \quad G''(\omega, \varepsilon) \geq 0. \tag{32}$$

Finally, at long times, the behavior of the relaxation modulus must be consistent with the conventional limits in the respective states. In the pre-gel states, the material is a viscoelastic liquid, and the stress must relax completely at infinite time. In the post-gel states, on the other hand, the percolated network must sustain a finite residual stress, causing the relaxation modulus to approach a nonzero equilibrium modulus. This constraint can be represented by:

$$\begin{gathered} G(t, \varepsilon) \to 0, t \to \infty \text{ (pre-gel)}, \\ G_r(t, \varepsilon) = G(t, \varepsilon) - G_e(\varepsilon) \to 0, t \to \infty \text{ (post-gel)}. \end{gathered} \tag{33}$$

These constraints, taken together, need to be satisfied by a physically admissible expression for relaxation modulus on either side of the critical gel point. Most of the fractional viscoelastic models that theorize the sol-gel transition follow the above-mentioned constraints. These constraints, however, by themselves do not specify how the model parameters evolve with $\varepsilon$ on the pre-gel as well as the post-gel side. Therefore, by using linear viscoelastic relations, the parameters are needed to be related to $\varepsilon$ as well as scaling exponents to validate various expressions mentioned in Table 1.

In addition, another important constraint needs to be validated to complete the linear viscoelastic expressions in the pre-gel and post-gel states. As the critical gel state is approached from either side, the logarithmic derivative of the relaxation modulus with respect to $\varepsilon$ or $p$ (note that $\varepsilon = p_c - p$ in the pre-gel region while $\varepsilon = p - p_c$ in the post-gel region) must exhibit a power-law dependence on time governed by the relaxation scaling exponent $\kappa$. In the pre-gel state, the expression is given by Eq. (14), while in the post-gel state it is given by Eq. (15). The negative sign in the expression for pre-gel state (Eq. (14)) is because, owing to $\varepsilon = p_c - p$, an increase in $\varepsilon$ corresponds to a departure from the gel point in the direction where $p$ decreases, which causes the modulus to decrease. Note that the Fourier transform of the scaling constraint given by Eq. (14) and (15) result in equivalent expressions for the dynamic moduli in the frequency domain given by Eqs. (16) and (17), and have been experimentally validated for a wide range of gel-forming systems. The second kind of constraints is the continuity requirement at the critical gel point, which constitutes the most stringent physical condition. As the material passes through the critical gel state, whether during the sol-gel or the gel-sol transition, the dynamic moduli must not only be continuous just as functions of $p$, given by Eq. (20), but their derivatives with respect to $p$ must also be continuous across $p_c$ leading to Eq. (21). Both the above-mentioned conditions must hold for all frequencies $\omega$. The present model development is, therefore, motivated by the imperative that the fractional viscoelastic model on the pre-gel and the post-gel sides cannot be developed in isolation, but must be jointly constrained so that conditions mentioned above are satisfied.

**Pre-Gel State**

**Pre-Gel Model 1. Fractional Maxwell Model with Springpots in Series**

To model the pre-gel states, we begin by considering the most general configuration consisting of $N$ springpots and one dashpot connected in series, as shown in Figure 1. In the pre-gel states, since the rheological response is that of viscoelastic liquids, series combination becomes a natural choice for modeling. Let the springpots be labeled by the index $k = 1, 2, \ldots, N$, where the $k$th springpot is characterised by a quasi-property $V_k$ and fractional order $\alpha_k \in (0,1)$. Without loss of generality, the exponents $\alpha_k$ can be ordered as $0 < \alpha_1 < \alpha_2 < \cdots < \alpha_N < 1$. The constitutive equation for $k$th springpot is given by:

$$\sigma(t) = V_k\, D^{\alpha_k} \gamma_k(t), \qquad k = 1,2, \ldots, N, \tag{34}$$

where $D^{\alpha_k} = d^{\alpha_k} X / dt^{\alpha_k}$ denotes the Caputo fractional derivative of order $\alpha_k$. The constitutive equation for the dashpot is given by:

$$\sigma(t) = \eta(\varepsilon)\, \dot{\gamma}_\eta(t), \tag{35}$$

where the viscosity of the dashpot $\eta = \eta(\varepsilon)$ depends on $\varepsilon$.

For springpots connected in series, the stress is identical in all the elements $\left(\sigma(t) = \sigma_\eta(t) = \sigma_1(t) = \cdots = \sigma_N(t)\right)$, while the strains in individual elements add $\left(\gamma(t) = \gamma_\eta + \sum_{k=1}^{N} \gamma_k(t)\right)$ to give the total strain. The Laplace transform $\left(\tilde{f}(q) = \mathcal{L}\{f(t)\} = \int_0^\infty e^{-q}\, f(t) dt\right)$ of the Caputo derivative of order $\alpha$ with zero initial conditions maps to $q^\alpha$, given by $\mathcal{L}\{D^\alpha f(t)\} = q^\alpha \tilde{f}(q)$, where $\tilde{f}(q)$ is the Laplace transform of $f(t)$ (Hereafter, we use a tilde on top to show the Laplace transform of that variable; we employ $q$ as the Laplace domain variable). Consequently, the Laplace transform of strain in $k$th springpot leads to $\tilde{\gamma}_k(q) = \tilde{\sigma}(q)/(V_k q^{\alpha_k})$ and in the dashpot is $\tilde{\gamma}_\eta(q) = \tilde{\sigma}(q)/(\eta q)$. $A$dding the strain contributions results in the total strain in the Laplace domain given by:

$$\tilde{\gamma}(q) = \tilde{\sigma}(q) \sum_{k=1}^{N} \frac{1}{V_k q^{\alpha_k}} + \frac{1}{\eta(\varepsilon) q}. \tag{36}$$

For the step-strain input: $\gamma(t) = \gamma_0 H(t)$, where $H(t)$ is the Heaviside unit step function and $\gamma_0$ is the magnitude of strain. Laplace transform of the step strain yields $q\tilde{\gamma}(q) = \gamma_0$ and using $\tilde{\sigma}(q) = \gamma_0 \tilde{G}(q)$, the expression of relaxation modulus in the Laplace domain is given by:

$$\tilde{G}(q) = \frac{1}{\sum_{k=1}^{N} \frac{q^{1-\alpha_k}}{V_k} + \frac{1}{\eta(\varepsilon)}}. \tag{37}$$

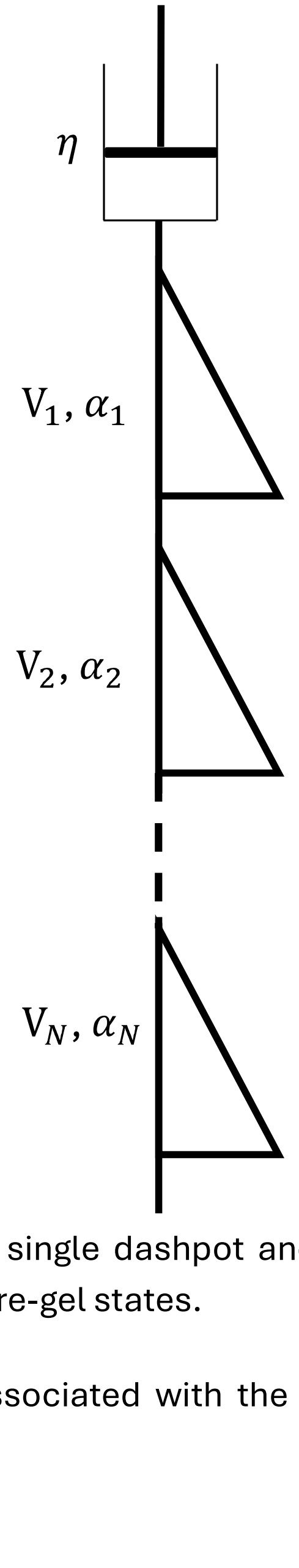

**Figure 1.** A series assembly of a single dashpot and $N$ springpots. It is a proposed fractional viscoelastic model for pre-gel states.

Factoring out the contribution associated with the smallest fractional exponent $\alpha_1$ leads to:

$$\tilde{G}(q,\varepsilon) = \frac{V_1\, q^{\alpha_1 - 1}}{1 + \sum_{k=2}^{N} r_k(\varepsilon) q^{\alpha_1 - \alpha_k} + \mu(\varepsilon)\, q^{\alpha_1 - 1}}, \tag{38}$$

where $r_k(\varepsilon) = V_1/V_k(\varepsilon)$ and $\mu(\varepsilon) = V_1/\eta(\varepsilon)$ are the ratios that depend on $\varepsilon$.

The dominant short-time term associated with $\tilde{G}(q)$ can be obtained by taking a limit of $q \to \infty$, which corresponds to $t \to 0^+$. For a material in the pre-gel states, therefore, $\lim_{q\to\infty} \tilde{G}(q)$ is expected to result in the Laplace transform of the Winter–Chambon form $\mathcal{L}\{G(t) \sim t^{-n}\} \to \tilde{G}(q) \sim q^{n-1}$. It can be seen that for Eq. (38) to reduce to Winter–Chambon form in a limit $q \to \infty$, we must satisfy $\alpha_1 = n < 1$ and $1 > \alpha_k > n$ for all $k \geq 2$. Furthermore, at the critical gel point ($\varepsilon \to 0$), the relaxation modulus must recover the Winter–Chambon form $G(t, \varepsilon = 0) = St^{-n}$. Since $\mathcal{L}\{St^{-n}\} = S\Gamma(1 - n)q^{n-1}$, that necessitates $r_k \to 0$ for $k \geq 2$ and $\mu \to 0$ in a limit $\varepsilon \to 0$, in addition to $V_1 = S\Gamma(1-n)$. This constraint suggests that: $r_k(\varepsilon) = c_k \varepsilon^{p_k}$ with $c_k > 0$ and $p_k > 0$, and hence all springpots for $k \geq 2$ become asymptotically negligible as the gel point is approached. Comparing $\eta_0 = \int_0^\infty G(t,\varepsilon)\, dt$ with $\tilde{G}(q,\varepsilon) = \int_0^\infty G(t,\varepsilon) e^{-qt}\, dt$, the zero-shear viscosity for the entire assembly is given by:

$$\eta_0 = \tilde{G}(q \to 0^+, \varepsilon) = \lim_{q \to 0^+} \frac{V_1\, q^{n-1}}{\mu(\varepsilon)\, q^{n-1}} = \frac{V_1}{\mu(\varepsilon)} \tag{39}$$

Eq. (39) suggests that the zero-shear viscosity is simply the dashpot viscosity for all $\varepsilon > 0$ without any condition on the remaining model parameters. We, therefore, set $\eta(\varepsilon) = \eta_S\, \varepsilon^{-s}$ with $\eta_* > 0$ to be an $\varepsilon$-independent material constant that leads to pre-gel scaling $\eta_0 \sim \varepsilon^{-s}$ given by Eq. (6). Through, Eq. (10), the power-law scaling exponent is given by: $s = (1-n)/\kappa_S$.

In the pre-gel states, a vital constraint that the expression of relaxation modulus must satisfy is given by Eq. (14) that requires the logarithmic derivative of the relaxation modulus with respect to $\varepsilon$, evaluated at $\varepsilon \to 0$, must exhibit a single power-law dependence on time. In order to impose the same, we expand Eq. (38) in a limit of $\varepsilon \to 0$ (or $r_k \to 0$ and $\mu \to 0$) leading to:

$$\tilde{G}(s,\varepsilon) \approx V_1\, q^{n-1} \left(1 - \sum_{k=2}^{N} r_k\,(\varepsilon)\, q^{n-\alpha_k} - \mu(\varepsilon)\, q^{n-1}\right) + O(\varepsilon^2). \tag{40}$$

Taking the term-by-term inverse Laplace transform of Eq. (40) leads to:

$$G(t,\varepsilon) = St^{-n}\left[1 - \sum_{k=2}^{N} \frac{\Gamma(1-n) r_k(\varepsilon)}{\Gamma(1+\alpha_k - 2n)}\, t^{\alpha_k - n} - \frac{\Gamma(1-n)}{\Gamma(2-2n)}\, \mu(\varepsilon)\, t^{1-n}\right] + O(\varepsilon^2). \tag{41}$$

Differentiating logarithm of Eq. (41) with respect to $\varepsilon$ and taking of $\varepsilon \to 0$ limit results in:

$$\left(\frac{\partial \ln G(t,\varepsilon)}{\partial \varepsilon}\right)_{\varepsilon\to 0} = -\sum_{k=2}^{N} \frac{\Gamma(1-n)}{\Gamma(1+\alpha_k-2n)}\left(\frac{dr_k}{d\varepsilon}\right)_{\varepsilon\to 0} t^{\alpha_k-n} - \frac{\Gamma(1-n)}{\Gamma(2-2n)}\left(\frac{d\mu}{d\varepsilon}\right)_{\varepsilon\to 0} t^{1-n}. \quad (42)$$

When a series $\sum_j A_j t^{\beta_j}$ is identically equal to $-B_S t^{\kappa_S}$, it is essential that one exponent must exactly equal $\kappa_S$ and has a coefficient $-B_S$; and all the other coefficients must vanish. For the dashpot, since $\mu(\varepsilon) = V_1/\eta(\varepsilon) \sim \varepsilon^s$, we get $d\mu/d\varepsilon \sim \varepsilon^{s-1}$. Therefore, for the second term of Eq. (42) to vanish in a limit of $\varepsilon \to 0$, we must have $s > 1$, so the dashpot term does not contribute to the logarithmic derivative constraint given by Eq. (14). Regarding the springpots, since the contribution from $k$th springpot is $\left(\frac{dr_k}{d\varepsilon}\right) t^{\alpha_k-n} = c_k p_k \varepsilon^{p_k-1} t^{\alpha_k-n}$, the terms with $p_k > p_{min}$ must vanish as $\varepsilon \to 0$. In addition, among all the springpots, only those that possess identical exponent $\alpha_k - n = \kappa_S$ contribute at leading order in $\varepsilon$. Other springpots that have different fractional orders necessarily vanish before $\alpha_k - n = \kappa_S$ term in the asymptotic limit $\varepsilon \to 0$. Furthermore, all springpots that possess the same order $\alpha_k = n + \kappa_S$ can be combined into a single effective springpot because in the series combination, their compliances simply add. Consequently, irrespective of the number of springpots initially introduced, the leading-order pre-gel response in the asymptotic limit $\varepsilon \to 0$ necessarily leads to an effective three-element model: two springpots and a dashpot in series.

For an assembly of $N$ springpots and a dashpot in series, the expression of relaxation modulus in the Laplace domain is given by Eq. (38) and can be represented as: $\tilde{G}(q,\varepsilon) = V_1\, q^{n-1}/(1+\mathcal{D}(q,\varepsilon))$. Since $\mathcal{D}(q,\varepsilon) \to 0$ as $\varepsilon \to 0$, we can express: $\tilde{G}(q,\varepsilon) = V_1 q^{n-1} \sum_{m=0}^{\infty} (-\mathcal{D}\,(q,\varepsilon))^m$, which can be further expanded in such a format so as to facilitate its analytical Laplace inverse. However, since $\mathcal{D}(q,\varepsilon)$ is a sum of several terms, getting analytical expression for relaxation modulus becomes mathematically cumbersome and may involve many approximations. Furthermore, the assembly of $N$ springpots and a dashpot in series has a major constraint as a model for the pre-gel states, as it is only applicable for $s > 1$. Since $s = (1-n)/\kappa_S$, the model can consider only those values of $n$ and $\kappa_S$ that follow: $\kappa_S + n < 1$. Considering both these limitations, and the fact that two springpots and a dashpot in series inevitably result in the leading-order pre-gel response in the asymptotic limit of $\varepsilon \to 0$, we restrict the model to this effective three-element assembly.

Based on this discussion, the first springpot has quasi-property $V_1 = S\Gamma(1-n)$ and fractional order $n$, while the second springpot has a quasi-property $V_2(\varepsilon)$ and fractional order $\alpha_2 = n + \kappa_2 \in (n, 1)$, where $\kappa_2$ is determined below. The dashpot has viscosity $\eta(\varepsilon) = \eta_S\, \varepsilon^{-s}$. The relaxation modulus in the Laplace domain given by Eq. (38) reduces to:

$$\tilde{G}(q,\varepsilon) = \frac{V_1\, q^{n-1}}{1 + r(\varepsilon) q^{-\kappa_2} + \mu(\varepsilon)\, q^{n-1}} = \frac{V_1\, q^{n-1}}{D(q,\varepsilon)}, \quad (43)$$

where $r(\varepsilon) = V_1/V_2(\varepsilon)$ and $\mu(\varepsilon) = V_1/\eta(\varepsilon)$. In Eq. (43), the quantities $r(\varepsilon)$ and $\mu(\varepsilon)$ are still unknown functions of $\varepsilon$. Instead of guessing their form, we require the model to obey the scaling form of Eq. (12) with the single diverging time given by Eq.(8), $\tau_{\max,S} = \tau_S \varepsilon^{-1/\kappa_S}$. Throughout, large $q$ corresponds heuristically to short times and small $q$ to long times ($q \sim 1/t$). However, the exact expressions below use the inverse transform. Laplace transform of Eq. (12) is given by: $\tilde{G}(q,\varepsilon) = \tau_{\max,S}^{1-n} F(q\tau_{\max,S})$, which suggests that $\varepsilon$ and $q$ may enter only through the product $q\tau_{\max,S}$. Numerator of Eq. (43) validates the same well. Rewriting the denominator of Eq. (43) as $D = 1 + r(\varepsilon)\tau_{\max,S}^{\kappa_2}(q\tau_{\max,S})^{-\kappa_2} + \mu(\varepsilon)\tau_{\max,S}^{1-n}(q\tau_{\max,S})^{n-1}$. It depends on $q\tau_{\max,S}$ alone only if $r(\varepsilon)\tau_{\max,S}^{\kappa_2}$ and $\mu(\varepsilon)\tau_{\max,S}^{1-n}$ are constants. Consequently, $r(\varepsilon) = \vartheta\, \tau_{\max,S}^{-\kappa_2}$ and $\mu(\varepsilon) = b\, \tau_{\max,S}^{n-1}$, where $\vartheta$ and $b$ are dimensionless, $\varepsilon$-independent constants. At first order in $\varepsilon$ the second springpot gives $G \simeq S\, t^{-n}\, [\, 1 - c_1\, r(\varepsilon)\, t^{\kappa_2}]$ with $c_1 = \Gamma(1-n)/\Gamma(1-n+\kappa_2)$. Since $\tau_{\max,S} = \tau_S \varepsilon^{-1/\kappa_S}$, we have $r \sim \varepsilon^{\kappa_2/\kappa_S}$. Consequently, $\partial \ln G/\partial\varepsilon$ at $\varepsilon \to 0$ is finite, nonzero and $\sim t^{\kappa_S}$, as per Eq. (14) demands, only if $\kappa_2 = \kappa_S$. This suggests $r = \vartheta_\varepsilon \varepsilon$ is linear in $\varepsilon$ with $\vartheta_\varepsilon = \vartheta \tau_S^{-\kappa_S}$. Since only this product is observable, we choose $\tau_S$ such that $\vartheta = 1$. Likewise, $\mu = \mu_0 \varepsilon^s$ with $\mu_0 = b\, \tau_S^{n-1} = S\Gamma(1-n)/\eta_S$, so $b = S\Gamma(1-n)\tau_S^{1-n}/\eta_S$ is a shape parameter carrying the viscosity prefactor $\eta_S$. The scaling form reproduces $\eta \propto \varepsilon^{-s}$ with $s = (1-n)/\kappa_S$, consistent with Eqs. (6) and (10). Hence, $D(q,\varepsilon) = 1 + \vartheta_\varepsilon\, \varepsilon q^{-\kappa_S} + \mu_0 \varepsilon^s q^{\,n-1}$. At short times (high $q$) both extra terms are small, $D \to 1$ and $\tilde{G} \to V_1 q^{n-1}$, which is the expression for the critical gel. Under constant stress, the compliances of elements in series add: $J(t) = t^n/[V_1\Gamma(1+n)] + t^{n+\kappa_S}/[V_2\Gamma(1+n+\kappa_S)] + t/\eta$. For $t \ll \tau_{\max,S}$ the first springpot dominates. The dashpot, whose strain grows fastest since $n + \kappa_S < 1$, dominates for $t > b^{-1/(1-n-\kappa_S)}\tau_{\max,S}$. This is the terminal viscous flow regime associated with the sol state.

For $t \ll b^{-1/(1-n-\kappa_S)}\tau_{\max,S}$ since the contribution from the dashpot is negligible, dropping the dashpot term from the denominator leads to the leading-order Laplace transform expression given by:

$$\tilde{G}(q,\varepsilon) \approx \frac{V_1\, q^{n+\kappa_S-1}}{q^{\kappa_S} + \vartheta_\varepsilon\, \varepsilon}, \tag{44}$$

This expression of the relaxation modulus in the Laplace domain can be transformed back into the time domain via the standard inverse Laplace transform identity:

$$\mathcal{L}^{-1}\left(\frac{q^{\alpha-\beta}}{q^{\alpha} + \lambda}\right) = t^{\beta-1} E_{\alpha,\beta}(-\lambda t^{\alpha}), \tag{45}$$

that involves the two-parameter Mittag-Leffler function defined as [64]:

$$E_{\alpha,\beta}(z) = \sum_{k=0}^{\infty} \frac{z^k}{\Gamma(\alpha k + \beta)}. \tag{46}$$

This two-parameter Mittag-Leffler function reduces to an exponential function for $\alpha = \beta = 1$, leading to: $E_{1,1}(z) = e^z$. Comparison of Eqs. (44) and (45) leads to $\alpha = \kappa_S$, $\beta = 1 - n$, and $\lambda = \vartheta_\varepsilon\, \varepsilon$. Inverting term by term:

$$G(t,\varepsilon) \approx S\,\Gamma(1-n)t^{-n}E_{\kappa_S,\,1-n}(-\vartheta_\varepsilon\,\varepsilon\, t^{\kappa_S}) \quad \ldots. \qquad t \ll b^{-1/(1-n-\kappa_S)}\tau_{\max,S}. \tag{47}$$

This expression is identical to the form of relaxation modulus for the two-springpot in series model as in a limit of $\varepsilon \to 0$ and for $t \ll b^{-1/(1-n-\kappa_S)}\tau_{\max,S}$ viscosity of the dashpot diverges so significantly that it does not contribute to the relaxation of stress.

However, in Eq. (43) the expression for relaxation modulus over the entire time duration, can be obtained by expressing: $D = 1 + y$ with $y = \vartheta_\varepsilon \varepsilon q^{-\kappa_S} + \mu_0\varepsilon^s q^{n-1}$. At large $q$, $|y| < 1$, the geometric series $1/(1+y) = \sum_m (-y)^m$ applies. Writing $y = y_1 + y_2$, with $y_1 = \vartheta_\varepsilon \varepsilon q^{-\kappa_S}$ and $y_2 = \mu_0 \varepsilon^s q^{n-1}$, and using $(-y)^m = (-1)^m (y_1 + y_2)^m$ together with the binomial theorem (Eq. 3.1.1 of [65]), $(y_1+y_2)^m = \Sigma_{j=0}^{m}[m!/(j!\,(m-j)!)]y_1^{m-j}y_2^j$ the geometric series becomes $1/(1+y) = \Sigma_{m=0}^{\infty}\Sigma_{j=0}^{m}(-1)^m[m!/(j!\,(m-j)!)]y_1^{m-j}y_2^j$, where $y_1^{m-j}y_2^j = (\vartheta_\varepsilon\varepsilon)^{m-j}(\mu_0\varepsilon^s)^j q^{-\kappa_S(m-j)-(1-n)j}$. Substituting into Eq. (43) gives:

$$\tilde{G} = V_1 \sum_{m=0}^{\infty}\sum_{j=0}^{m}(-1)^m \frac{m!}{j!\,(m-j)!}(\vartheta_\varepsilon\varepsilon)^{m-j}(\mu_0\varepsilon^s)^j\; q^{-\varsigma_{mj}}, \tag{48}$$

where $\varsigma_{mj} = 1 - n + \kappa_S(m-j) + (1-n)j$. Each term is a power of $q$, and $\mathcal{L}^{-1}\{q^{-\varsigma}\} = t^{\varsigma-1}/\Gamma(\varsigma)$.Inverting term by term,

$$G(t,\varepsilon) = S\,\Gamma(1-n)\,t^{-n}\sum_{m=0}^{\infty}\sum_{j=0}^{m}\frac{(-1)^m m!}{j!\,(m-j)!}\frac{(\vartheta_\varepsilon\varepsilon t^{\kappa_S})^{m-j}\,(\mu_0\varepsilon^s t^{\,1-n})^j}{\Gamma(1-n+\kappa_S(m-j)+(1-n)j)}. \tag{49}$$

The geometric series needs $|y| < 1$, but the time-domain series converges for all $t$, because the Gamma functions in the denominator grow faster than any exponential. This is a bivariate Mittag-Leffler function [66]. In scaling form,

$$\begin{gathered} G(t,\varepsilon) = S\,\Gamma(1-n)\,t^{-n}\Phi\big(t/\tau_{\max,S};b\big) \\ \Phi(x;b) = \sum_{m=0}^{\infty}\sum_{j=0}^{m}\frac{(-1)^m m!}{j!\,(m-j)!}\frac{b^j\,x^{\kappa_S(m-j)+(1-n)j}}{\Gamma\big((1-n+\kappa_S(m-j)+(1-n)j)\big)} \end{gathered} \tag{50}$$

with $\Phi(0) = 1$ and $\Phi(\infty) = 0$, which is Eq. (12) without approximation. In Eq. (50), the terms with $j = 0$ sum to $E_{\kappa_S,1-n}(-\vartheta_\varepsilon\varepsilon t^{\kappa_S})$ which is exactly Eq. (47).

Furthermore, in Eq. (50), at a fixed $t$ and small $\varepsilon$, only the terms with $m \le 1$ survive leading to,

$$G(t,\varepsilon) \simeq S\,t^{-n}\left[1 - \frac{\Gamma(1-n)}{\Gamma(1-n+\kappa_S)}\,\vartheta_\varepsilon\,\varepsilon\,t^{\kappa_S} - \frac{\Gamma(1-n)}{\Gamma(2-2n)}\,\mu_0\,\varepsilon^s\,t^{1-n} + O(\varepsilon^2)\right]. \tag{51}$$

Differentiating the logarithm with respect to $\varepsilon$ and taking $\varepsilon \to 0$, the dashpot term gives $\propto s\,\varepsilon^{s-1}$, which vanishes because $s > 1$. Hence

$$\left(\frac{\partial \ln G(t,\varepsilon)}{\partial \varepsilon}\right)_{\varepsilon \to 0} = -\frac{\Gamma(1-n)}{\Gamma(1-n+\kappa_S)}\,\vartheta_\varepsilon\, t^{\kappa_S}. \quad (52)$$

Equating this to Eq. (14) uniquely gives $\vartheta_\varepsilon = B_S\,\Gamma(\kappa_S + 1 - n)/\Gamma(1-n)$.

Because $G(t) = 0$ for $t < 0$, the Fourier relation of Eq. (43) equals the Laplace transform evaluated at $q = i\omega$:

$$G^*(\omega,\varepsilon) = i\omega \int_0^\infty G\,(t,\varepsilon)\, e^{-i\omega t}\, dt = i\omega\, \tilde{G}(q = i\omega, \varepsilon). \quad (53)$$

Inserting Eq. (43) gives the exact result for the assembly, with no approximation:

$$G^*(\omega,\varepsilon) = \frac{S\Gamma(1-n)\,(i\omega)^n}{1 + \vartheta_\varepsilon \varepsilon\,(i\omega)^{-\kappa_S} + \mu_0 \varepsilon^s (i\omega)^{n-1}}. \quad (54)$$

At fixed $\omega$ and $\varepsilon \to 0$, the denominator tends to 1 and Eq. (54) reduces to the Winter–Chambon Eq. (2). At low frequency, it gives $G^* \to i\omega\eta_S \varepsilon^{-s}$, which is pure viscous flow. Resolving Eq. (54) into real and imaginary parts leads to:

$$G' = \frac{S\Gamma(1-n)\,\omega^n\left[D_r \cos\frac{\pi n}{2} - D_i \sin\frac{\pi n}{2}\right]}{D_r^2 + D_i^2}, \quad (55)$$

$$G'' = \frac{S\Gamma(1-n)\,\omega^n\left[D_r \sin\frac{\pi n}{2} + D_i \cos\frac{\pi n}{2}\right]}{D_r^2 + D_i^2}, \quad (56)$$

where

$$D_r = 1 + \vartheta_\varepsilon \varepsilon\,\omega^{-\kappa_S} \cos\frac{\pi\kappa_S}{2} + \mu_0 \varepsilon^s \omega^{n-1} \sin\frac{\pi n}{2}, \quad (57)$$

$$D_i = \vartheta_\varepsilon \varepsilon\,\omega^{-\kappa_S} \sin\frac{\pi\kappa_S}{2} + \mu_0 \varepsilon^s \omega^{n-1} \cos\frac{\pi n}{2}. \quad (58)$$

Differentiating Eqs. (55) and (56) and taking a limit $\varepsilon \to 0$, leads to:

$$\left.\frac{\partial G'}{\partial \varepsilon}\right|_{\varepsilon \to 0} = -S\Gamma(1-n)\,\vartheta_\varepsilon\,\omega^{n-\kappa_S} \cos\left(\frac{\pi(n-\kappa_S)}{2}\right), \quad (59)$$

$$\left.\frac{\partial G''}{\partial \varepsilon}\right|_{\varepsilon \to 0} = -S\Gamma(1-n)\,\vartheta_\varepsilon\,\omega^{n-\kappa_S} \sin\left(\frac{\pi(n-\kappa_S)}{2}\right). \quad (60)$$

Both limits are finite and independent of $\varepsilon$. Dividing Eqs. (59) and (60), respectively by the critical-gel values, Eqs. (3) and (4), lead to the corresponding derivatives in semilogarithmic form that scale as $\omega^{-\kappa_S}$, as required by Eq. (16). Their ratio is independent of frequency and equals $C_{S\to G} = (\partial \ln G'/\,\partial \ln G'')_{\varepsilon \to 0} = \cot[(n-\kappa_S)\pi/$ $2]\tan(n\pi/2)$, i.e. Eq. (24), with $(\partial G''/\,\partial G')_{\varepsilon\to 0} = \tan[(n-\kappa_S)\pi/2]$. This reproduces the model-agnostic result of Appendix A.

As discussed, the proposed expression of the relaxation modulus given by Eq. (51) is physically admissible only for $s > 1$ resulting in $n + \kappa_S < 1$ that also ensures complete monotonicity of the same. While this constraint does describe a meaningful class of gel-forming systems, it is indeed more restrictive than the full physically accessible parameter space. Particularly, systems with relatively large $n$ are thereby excluded from the admissible description. Furthermore, since the dashpot controls terminal flow and viscosity, the expression for relaxation modulus needs a double series or numerical inversion. We therefore turn to the three-parameter Mittag-Leffler (Prabhakar) function, which is admissible for all $0 < \kappa_S < n < 1$ and is a single series.

**Pre-Gel Model 2. Model based on Mittag-Leffler-Prabhakar function**

The restriction $n + \kappa_S < 1$ of Pre-Gel Model 1, together with the absence of a compact single-series form for its time-domain response, leads to the three-parameter Mittag–Leffler function proposed by Prabhakar [67]. This function generalizes the two-parameter Mittag–Leffler functions by introducing an additional parameter $\gamma$ and is defined as: [46, 64, 67, 68]

$$E_{\alpha,\beta}^{\gamma}(z) = \sum_{k=0}^{\infty} \frac{\Gamma(\gamma+k)}{k!\,\Gamma(\gamma)} \frac{z^k}{\Gamma(\alpha k+\beta)} \tag{61}$$

The parameter $\gamma$ appears through the factor, $\Gamma(\gamma+k)/\,k!\,\Gamma(\gamma)$, which systematically provides an additional degree of freedom by modulating each term in the series. Interestingly a specific quantity, obtained by multiplying the Prabhakar function by a power-law prefactor $t^{\beta-1}E_{\alpha,\beta}^{\gamma}(-\lambda t^{\alpha})$, has emerged as a vital building block in various branches of science and engineering, particularly in the theory of anomalous relaxation, dielectric response, and fractional differential equations [46, 64, 67-69]. Furthermore, the Fourier transform of the function bearing a form $t^{\beta-1}E_{\alpha,\beta}^{\gamma}(-\lambda t^{\alpha})$ results in Havriliak–Negami relaxation function that is routinely used in electromagnetism to account for both the asymmetrical shape and the broadness of the dielectric dispersion curve [51, 70, 71]. Considering mathematical equivalence of dielectric response as well as viscoelasticity [72] it is therefore, natural that the Prabhakar function is directly applicable to the latter field. One of the reasons for the Prabhakar function's significance is the remarkably compact form of its Laplace transform given by [46, 64, 67, 68]:

$$\mathcal{L}\left\{t^{\beta-1}E^{\gamma}_{\alpha,\beta}(-\lambda t^{\alpha})\right\}(q)=\frac{q^{\alpha\gamma-\beta}}{(q^{\alpha}+\lambda)^{\gamma}}. \tag{62}$$

Note that, for $\gamma = 1$, the three-parameter Mittag-Leffler-Prabhakar function reduces exactly to the standard two-parameter Mittag-Leffler given by Eq. (46). The conditions for which $t^{\beta-1}E^{\gamma}_{\alpha,\beta}(-x)$ with $x \geq 0$ is completely monotonic are given by: $0 < \alpha \leq 1$ and $0 < \gamma \leq \beta/\alpha$ [46, 64, 67, 68]. For the proposed relaxation modulus based on the Prabhakar function, since the primary basis is adapted from the dashpot-free limit of Pre-Gel Model 1 (Eq. (47)), a two-parameter Mittag-Leffler function given by Eq. (46) with $\alpha = \kappa_S$ and $\beta = 1 - n$, the condition on $\gamma$ for monotonicity becomes:

$$0<\gamma\leq\frac{1-n}{\kappa_S}. \tag{63}$$

For $\gamma = 1$, as expected, Eq. (63) exactly reduces to the constraint of the fractional Maxwell model (Pre-Gel Model 1) $n + \kappa_S \leq 1$ of the two-parameter Mittag-Leffler function, i.e. of the dashpot-free limit of Pre-Gel Model 1 given by Eq. (47). However, for $\gamma < 1$, the above constraint becomes $\kappa_S \leq (1-n)/\gamma$, and for any permissible value of $n$, limit on $\kappa_S$ gets progressively relaxed as $\gamma$ decreases further below the unity. Consequently, for any pair $(n, \kappa_S)$ with $n + \kappa_S > 1$, there always exists a sufficiently small $\gamma \in (0,1)$ such that the three-parameter Mittag-Leffler Prabhakar function $E^{\gamma}_{\kappa_S,1-n}(-\vartheta_\varepsilon \varepsilon t^{\kappa_S})$ is always monotonic. Therefore, the regularization induced by the parameter $\gamma$ extends the physically admissible parameter space to all the admissible values of $n \in (0,1)$ and $\kappa_S \in (0,1)$, while preserving every structural feature of the model (it should be noted here that according to Joshi [40], values of $n < \kappa_S$ are unphysical).

With this background, we replace the two-parameter Mittag-Leffler function of the dashpot-free limit of Pre-Gel Model 1, Eq. (47), by the three-parameter Mittag-Leffler-Prabhakar function that results in the generalized pre-gel relaxation modulus given by:

$$\begin{aligned}G(t,\varepsilon)&=S\Gamma(1-n)\,t^{-n}\,E^{\gamma_S}_{\kappa_S,\,1-n}(-\vartheta_\varepsilon \varepsilon t^{\kappa_S})\\&=S\Gamma(1-n)\sum_{k=0}^{\infty}\frac{\Gamma(\gamma_S+k)}{k!\,\Gamma(\gamma_S)}\frac{(-\vartheta_\varepsilon\varepsilon)^k}{\Gamma(\kappa_S k+1-n)}t^{\kappa_S k-n},\end{aligned} \tag{64}$$

which results in the generalized pre-gel relaxation modulus, which, unlike its dashpot-free origin (Eq. (47)), has a finite zero-shear viscosity. This expression (Eq (64)) of relaxation modulus satisfies all constraints of Table 1 for $0 < \gamma_S \leq (1-n)/\kappa_S$. The same construction was used by Joshi [40], for the conventional model. The above expression of the relaxation modulus can be generalized to incorporate multimodal representation given by:

$$G(t,\varepsilon) = S\Gamma(1-n)t^{-n}\sum_{j=1}^{P} w_j E_{\kappa_S,1-n}^{\gamma_S}\left(-\vartheta_{S,j}\varepsilon t^{\kappa_S}\right), \tag{65}$$

where the weights satisfy $0 \le w_j \le 1$ and $\sum_{j=1}^{P} w_j = 1$, and $\vartheta_{S,j}$ is the (mode-specific) pre-gel material constant associated with the $j$th Prabhakar contribution. Incorporating the expression for Prabhakar function given by Eq. (61) for the single-mode in Eq. (65) and changing the order of summation leads to the relaxation modulus given by:

$$\begin{aligned} G(t,\varepsilon) &= S\,t^{-n}\Gamma(1-n)\sum_{k=0}^{\infty}\frac{\Gamma(\gamma_S+k)\overline{\vartheta_S^k}(-\varepsilon)^k}{k!\,\Gamma(\gamma_S)\Gamma(k\kappa_S+1-n)}t^{k\kappa_S} \\ &= S\,t^{-n}\left[1-\frac{\gamma_S\Gamma(1-n)}{\Gamma(\kappa_S+1-n)}\overline{\vartheta_S}\,\varepsilon\,t^{\kappa_S}+\mathcal{O}(\varepsilon^2)\right], \end{aligned} \tag{66}$$

where $\overline{\vartheta_S^k}$ is given by

$$\overline{\vartheta_S^k} = \sum_{j=1}^{P} w_j \vartheta_{S,j}^k. \tag{67}$$

The second term of Eq. (66) shows the expression of relaxation modulus in the series to first order in $\varepsilon$.

Differentiating the logarithm of $G(t,\varepsilon)$ with respect to $\varepsilon$ and taking the limit $\varepsilon \to 0$ gives:

$$\left(\frac{\partial \ln G(t,\varepsilon)}{\partial \varepsilon}\right)_{\varepsilon\to 0} = -\frac{\gamma_S\,\Gamma(1-n)}{\Gamma(\kappa_S+1-n)}\,\overline{\vartheta_S^1}\,t^{\kappa_S}. \tag{68}$$

Equating this expression to the constraint given by Eq. (14) from Table 1 leads to a unique relationship between $\overline{\vartheta_S^1}$ and $B_s$ given by:

$$\gamma_S\overline{\vartheta_S^1} = B_s\,\frac{\Gamma(\kappa_S+1-n)}{\Gamma(1-n)} \tag{69}$$

This is the multimodal Prabhakar analog of the single-mode relation mentioned in the previous section. In Eq. (69), the constraint couples $\gamma$ parameter to $\overline{\vartheta_S^1}$, which is the weighted average of $\vartheta_{S,j}$. As a result, the leading-order departure from the critical gel state is entirely governed by $\gamma_S\overline{\vartheta_S^1}$. For a single mode $P=1$, $\overline{\vartheta_S^1}$ reduces to $\vartheta_\varepsilon$. For $\gamma_S=1$, the two-parameter Mittag-Leffler function, i.e. the dashpot-free limit of Pre-Gel Model 1, the first-order relation $\vartheta_\varepsilon = B_S \Gamma(\kappa_S + 1 - n)/\Gamma(1-n)$ of the previous section is recovered exactly.

Next, we obtain the longest relaxation time and the zero-shear viscosity associated with the relaxation modulus function given by Eq. (66). In this multimodal representation, each $j$th mode possesses its own characteristic crossover timescale, which is obtained by setting $\vartheta_{S,j}\varepsilon t^{\kappa_S} = \left(t/\tau_{max,S,j}\right)^{\kappa_S}$ in the argument of the $j$th term in the Prabhakar function: $\tau_{max,S,j} = \left(\vartheta_{S,j}\,\varepsilon\right)^{-1/\kappa_S} = \tau_{S,j}\,\varepsilon^{-1/\kappa_S}$, where $\tau_{S,j} = \vartheta_{S,j}^{-1/\kappa_S}$ is the coefficient that is independent of $\varepsilon$. It can be seen that, all modes share the identical power-law coefficient $-1/\kappa_S$, and the overall longest relaxation time of the system is that of the slowest mode given by largest $\tau_{S,j}$ or smallest value of $\vartheta_{S,j}$. If we denote that coefficient as $\tau_S = \max_j(\vartheta_{S,j}^{-1/\kappa_S})$, the overall longest relaxation time can be expressed as:

$$\tau_{max,S} = \tau_S\,\varepsilon^{-1/\kappa_S}. \tag{70}$$

The divergence of $\tau_{max,S}$ as $\varepsilon \to 0$ therefore gets preserved in the multimodal form with the same exponent $1/\kappa_S$ as in the single-mode case.

In order to obtain the zero-shear viscosity, rather than using the standard integral expression used in the previous section, we make use of the Laplace transform approach. The Laplace transform of the multimodal relaxation modulus expressed by Eq. (65) is given by:

$$\tilde{G}(q,\varepsilon) = S\,\Gamma(1-n)\sum_{j=1}^{P} w_j \frac{q^{\,n-1+\kappa_S\gamma_S}}{\left(q^{\kappa_S}+\vartheta_{S,j}\,\varepsilon\right)^{\gamma_S}}. \tag{71}$$

As discussed before, the zero-shear viscosity is given by $\eta_0 = \tilde{G}(0)$. As $q \to 0^+$, the denominator of each term in Eq. (71) approaches $\left(\vartheta_{S,j}\,\varepsilon\right)^{\gamma_S}$, which is a finite constant for $\varepsilon > 0$. The numerator of Eq. (71), however, remains unchanged and remains $q^{\gamma_S\kappa_S-(1-n)}$. Under these conditions, for $\eta_0$ to be finite and to diverge with the correct power-law as $\varepsilon \to 0$, the numerator, therefore, must reduce to unity in the limit $q \to 0^+$. This physical constraint requires $\gamma_S\kappa_S = 1-n$, leading to unique determination of $\gamma_S$ given by: $\gamma_S = (1-n)/\kappa_S$, which interestingly also fulfils the monotonicity condition given by $\gamma_S \le (1-n)/\kappa_S$. The knowledge of Prabhakar exponent $\gamma_S$, which is independent of the number of modes, facilitates estimation of the zero-shear viscosity given by:

$$\eta_0 = S\,\Gamma(1-n)\mathcal{A}_{n,\kappa_S}\varepsilon^{-(1-n)/\kappa_S} = S\,\Gamma(1-n)\mathcal{A}_{n,\kappa_S}\varepsilon^{-\gamma_S}, \tag{72}$$

where the $\varepsilon$-independent prefactor $\mathcal{A}_{n,\kappa_S}$ is given by $\mathcal{A}_{n,\kappa_S} = \sum_{j=1}^{P} w_j \vartheta_{S,j}^{-(1-n)/\kappa_S} > 0$, which depends only on the mode parameters and critical exponents. Expression of viscosity given by Eq. (72) is a profound result, as it leads to the established expression for the power-law divergence of viscosity given by $\eta_0 \sim \varepsilon^{-s}$, which on one hand validates the scaling relation $s = (1-n)/\kappa_S$, but on the other hand, uniquely determines the Prabhakar exponent $\gamma_S$ to be:

$$\gamma_S = \frac{1-n}{\kappa_S} = s. \tag{73}$$

Similarly, Eq. (72) also suggests: $\eta_S = S\,\Gamma(1-n)\mathcal{A}_{n,\kappa_S}$.

The fact that analytically the Prabhakar exponent $\gamma_S$ is exactly equal to the viscosity divergence exponent $s$ has profound significance. Its value is fixed by $n$ and $\kappa$, and it can be compared with independently measured viscosity data through Eq. (73). In the framework of percolation theory, in the pre-gel state, it is the divergence in the size of the largest finite clusters that governs the divergence of the zero-shear viscosity as the system approaches the gel point. Mathematically, $\gamma_S$ governs the modal weight coefficient, $W_k(\gamma_S) = \Gamma(\gamma_S + k)/[k!\,\Gamma(\gamma_S)]$, which alters the relative contribution of higher-order, long-time relaxation modes ($k \geq 1$) compared to the uniform weighting ($\gamma_S = 1$) of the two-parameter Mittag-Leffler response. For $\gamma_S > 1$, these higher-order modes get amplified, causing broadening of the relaxation spectrum that may reflect a highly polydisperse sol composed of complex fractal clusters. On the other hand, $\gamma_S < 1$ dampens higher $k$ modes, suggesting narrowing of the relaxation spectrum compared to that associated with $\gamma_S = 1$ and suggesting presence of more structurally homogeneous cluster size distribution. Therefore, it is not surprising that an increasing value of $\gamma_S$ or s suggests a broadening of the relaxation spectra that leads to a sharper divergence of viscosity.

As discussed before, the Fourier transform of the Prabhakar function, on which the multimodal pre-gel relaxation modulus (Eq. (65)) is based, leads to the Havriliak–Negami relaxation function. In the following discussion, we make this identification explicit since the Havriliak–Negami form is ubiquitous in dielectric spectroscopy [67, 70, 71]. Substituting the relation $\gamma_S \kappa_S = 1-n$ in the Laplace-transformed relaxation modulus given by Eq. (71), results in a compact form given by:

$$\tilde{G}(q,\varepsilon) = S\,\Gamma(1-n) \sum_{j=1}^{P} \frac{w_j}{\left(q^{\kappa_S} + \vartheta_{S,j}\varepsilon\right)^{\gamma_S}}. \tag{74}$$

Using the mode-specific crossover time, $\tau_{max,S,j} = (\vartheta_{S,j}\varepsilon)^{-1/\kappa_S}$, and factoring $\vartheta_{S,j}\varepsilon = \tau_{max,S,j}^{-\kappa_S}$ out of the bracket gives:

$$\tilde{G}(q,\varepsilon) = S\,\Gamma(1-n)\sum_{j=1}^{P} w_j\ \tau_{max,S,j}^{1-n}\left[1+\left(q\tau_{max,S,j}\right)^{\kappa_S}\right]^{-\gamma_S}, \tag{75}$$

where the bracketed quantity, $[1+(q\tau)^{\alpha_{HN}}]^{-\gamma_{HN}}$, is the canonical, dimensionless Havriliak–Negami relaxation function of dielectric spectroscopy [67, 70, 71], with $\alpha_{HN} = \kappa_S$ and $\gamma_{HN} = \gamma_S$. Eq. (75) directly leads to the complex modulus in exactly the same closed Havriliak–Negami form given by,

$$G^*(\omega,\varepsilon) = S\,\Gamma(1-n)\,i\omega\sum_{j=1}^{P} w_j\ \tau_{max,S,j}^{1-n}\left[1+\left(i\omega\,\tau_{max,S,j}\right)^{\kappa_S}\right]^{-\gamma_S}, \tag{76}$$

The condition $\gamma_S = (1-n)/\kappa_S$ facilitates mapping of the Prabhakar function (Eq. (62)) onto textbook Havriliak–Negami function used throughout the dielectric literature [67, 70, 71, 73]. In addition, the above relationship: $\gamma_S = s$, corresponds to the marginal (boundary) case of the complete-monotonicity condition (Eq. (63)), discussed by Mainardi and Garrappa [68] as the boundary of admissible non-Debye relaxation. The Havriliak–Negami form was introduced for both dielectric and mechanical relaxation [67, 70, 71, 73], and Prabhakar-type kernels have been used in fractional viscoelasticity [74]. What is new here is that the Havriliak–Negami form emerges as the exact pre-gel complex modulus of a model constructed to satisfy the gelation scaling relations.

In dielectric spectroscopy, the Havriliak–Negami function parameters $\alpha_{HN}$ and $\gamma_{HN}$ govern the asymmetry and broadness of the spectra. More specifically, $\alpha_{HN}$ controls the symmetric broadening of the loss peak about the mean relaxation time. On the other hand, $\gamma_{HN}$ dictates its high-frequency asymmetry (skewness) [67, 70, 71].Both parameters are usually obtained by fitting experimental data. However, for the present pre-gel model these parameters are directly associated with characteristics of the sol–gel transition. Both $\alpha_{HN} = \kappa_S$ and $\gamma_{HN} = \gamma_S = s$ are related, respectively, to the divergence of the longest relaxation time ($\tau_{max,S} \sim \varepsilon^{-1/\kappa_S}$, Eq. (8)) and viscosity ($\eta_0 \sim \varepsilon^{-s}$, $s = (1-n)/\kappa_S$ Eq. (6)). Physically, $\alpha_{HN} = \kappa_S$ sets the self-similar, fractal broadening of the relaxation-time spectrum generated by the polydisperse population of finite clusters as the percolation threshold is approached, which is governed by the single correlation length $\xi(\varepsilon) \sim \varepsilon^{-\nu}$ [5, 75, 76]. As discussed before, the parameter $\gamma_S$, through the weighting factor $W_k^{(\gamma_S)} = \Gamma(\gamma_S + k)/[k!\ \Gamma(\gamma_S)]$, governs how the slower relaxation modes related to larger finite clusters are weighted relative to the faster modes. In the dielectric literature this asymmetry parameter $\gamma_{HN} = \gamma_S = s$ usually suggests intensity of cooperative or spatially heterogeneous relaxation [73, 77]. In the present scenario, that heterogeneity can be associated with the polydispersity of the pre-gel cluster-size distribution, while the value of $\gamma_{HN} = \gamma_S$ is fixed by the viscosity exponent $s$, which can be measured independently. The cluster-size distribution itself is not derived here. The two limiting cases of Havriliak–Negami function are as follows. $\gamma_S = 1$ reduces the model to the Cole–Cole form [51] leading to the two parameter Mittag Leffler model (Eq. (47)) that possesses symmetric spectral broadening with uniform modal weights, corresponding to the dashpot-free limit of Pre-Gel Model 1. $\kappa_S = 1$ corresponds to the Davidson–Cole form [51] resulting in a purely asymmetric long-time tail. However, for the gelation dynamics, the value of $\kappa_S$ is in the vicinity of 0.2 and hence that limit is never realized. This proposed equivalence is a specific realization of the correspondence between rheological and dielectric response discussed by Mittal et al. [72]. Remarkably, the present work substantiates this correspondence with a testable identity.

Expanding the expression given by Eq. (76), with $\tau_{max,S,j} = (\vartheta_{S,j}\varepsilon)^{-1/\kappa_S}$, in ascending powers of $\vartheta_{S,j}\varepsilon(i\omega)^{-\kappa_S}$ leads to more convenient form of the complex modulus (which can be easily resolved in real and imaginary parts) given by:

$$G^*(\omega,\varepsilon) = S\,\Gamma(1-n)\sum_{k=0}^{\infty}\frac{\Gamma(\gamma_S+k)}{\Gamma(\gamma_S)}\frac{\overline{\vartheta_S^k}(-\varepsilon)^k}{k!}(i\omega)^{n-k\kappa_S}. \tag{77}$$

This form can be obtained by taking term-by-term Fourier transform of relaxation modulus (Eq. (66)). Since for $k = 0$, $\overline{\vartheta_S^0} = \sum_{j=1}^{P} w_j = 1$, the $k = 0$ term always recovers the Winter–Chambon critical gel expression given by Eq. (2). The multimodal structure, therefore, enters the complex modulus through successive moments of the distribution of $\vartheta_{S,j}$. Resolving $G^*$ into its real and imaginary parts, the storage and loss moduli are given by:

$$G'(\omega,\varepsilon) = S\,\Gamma(1-n)\sum_{k=0}^{\infty}\frac{\Gamma(\gamma_S+k)}{\Gamma(\gamma_S)}\frac{\overline{\vartheta_S^k}(-\varepsilon)^k}{k!}\omega^{n-k\kappa_S}\cos\left(\frac{(n-k\kappa_S)\pi}{2}\right) \tag{78}$$

$$G''(\omega,\varepsilon) = S\,\Gamma(1-n)\sum_{k=0}^{\infty}\frac{\Gamma(\gamma_S+k)}{\Gamma(\gamma_S)}\frac{\overline{\vartheta_S^k}(-\varepsilon)^k}{k!}\omega^{n-k\kappa_S}\sin\left(\frac{(n-k\kappa_S)\pi}{2}\right) \tag{79}$$

In the limit $\varepsilon \to 0$, only the $k = 0$ term survives in each series, and both moduli recover the Winter–Chambon critical gel expressions given by Eqs. (3) and (4).

We now differentiate $G'$ and $G''$ with respect to $\varepsilon$ and take the limit $\varepsilon \to 0$. In this limit, only the $k = 1$ term in each series survives ($k = 0$ term is independent of $\varepsilon$ and its derivative vanishes; while the terms with $k \geq 2$ tend to zero in the limit $\varepsilon \to 0$), leading to:

$$\left.\frac{\partial G'}{\partial \varepsilon}\right|_{\varepsilon\to 0} = -S\,\Gamma(1-n)\gamma_S\overline{\vartheta_S^1}\omega^{n-\kappa_S}\cos\left(\frac{(n-\kappa_S)\pi}{2}\right) \tag{80}$$

$$\left.\frac{\partial G''}{\partial \varepsilon}\right|_{\varepsilon\to 0} = -S\,\Gamma(1-n)\gamma_S\overline{\vartheta_S^1}\omega^{n-\kappa_S}\sin\left(\frac{(n-\kappa_S)\pi}{2}\right) \tag{81}$$

It can be seen that both derivatives are finite and independent of $\varepsilon$. Dividing the above derivatives by the respective critical gel moduli at $\varepsilon = 0$, and using Eq. (69) we get the logarithmic derivatives that show an expected $\omega^{-\kappa_S}$ scaling. The ratio of these two logarithmic derivatives is frequency-independent and is equal to $C_{S\to G}$, whose model-agnostic expression is given by Eq. (A15) in Appendix A and Eq. (24) in Table 1.

A comparative assessment of the two pre-gel models developed above leads to a conceptually profound issue. Pre-Gel Model 1 is constructed as a series assembly of two springpots and a dashpot, which captures viscosity divergence mechanically through the dashpot viscosity $\eta(\varepsilon) = \eta_S\varepsilon^{-s}$. Furthermore, the exact response of this assembly is given by the complex modulus Eq. (54) and the bivariate Mittag-Leffler series Eq. (50). However, as a major drawback, this model is restricted to $n + \kappa_S < 1$. We also solve the dashpot-free limit (consisting of only two springpots in series) resulting in the two-parameter Mittag-Leffler function (Eq. (47)), which is compact but has no finite viscosity. On the other hand, Pre-Gel Model 2, invokes the Prabhakar function, which is a three-parameter generalization of the two-parameter Mittag-Leffler function. It is admissible for all $0 < \kappa_S < n < 1$, and has a finite zero-shear viscosity that diverges with the correct power-law as shown by Eq. (72) with Prabhakar parameter $\gamma_S$ uniquely taking the value of the viscosity divergence exponent: $\gamma_S = s$. This observation suggests that the conventional fractional viscoelastic approach for modelling the pre-gel states may not be universally applicable since it excludes gels with $n + \kappa_S \geq 1$. The present work essentially shows that the relaxation modulus based on the Prabhakar function, which does not have a fractional viscoelastic analog, may represent a more natural and complete description of the relaxation dynamics of the pre-gel state approaching the critical gel state.

## Post-Gel State

### Post-Gel Model 1. A Parallel Fractional Network of a Hookean Spring and an Array of Springpots

The critical gel state is characterized by the weakest space-spanning percolated network. As a system transitions into the post-gel state, the emerging percolated network gets progressively denser and capable of sustaining a finite stress in the limit of infinite time. The corresponding rheological signature is that the relaxation modulus no longer decays to zero at long times but instead approaches a non-zero equilibrium modulus $G_e(\varepsilon)$ that grows with $\varepsilon$. At the same time, on short and intermediate timescales, the material does retain the power-law viscoelastic character associated with the critical gel state, which must be recovered in the limit $\varepsilon \to 0$. These two coexisting but distinct features impose a natural constraint on the form of any admissible fractional viscoelastic model. Obviously, the elastic residual stress, which represents the permanent stress stored in the percolated network backbone, has to be associated with a purely elastic, non-relaxing element: a Hookean spring. The Winter–Chambon power-law expression that leads to the progressive stress relaxation, on the other hand, must be captured by one or more springpots, each of which relaxes completely at long times. Since in a parallel assembly the total stress is the sum of the stresses in each element having a common strain, a parallel arrangement of a Hookean spring with an array of springpots, therefore, becomes the most natural fractional representation for the post-gel states. This parallel arrangement is the analogue of the classical Kelvin–Voigt model that sets a non-zero equilibrium modulus through the spring and leads to a complete spectrum of power-law relaxations through the springpots. This also suggests that the parallel arrangement of one spring and at least one springpot is the minimum fractional assembly needed to satisfy both requirements simultaneously.

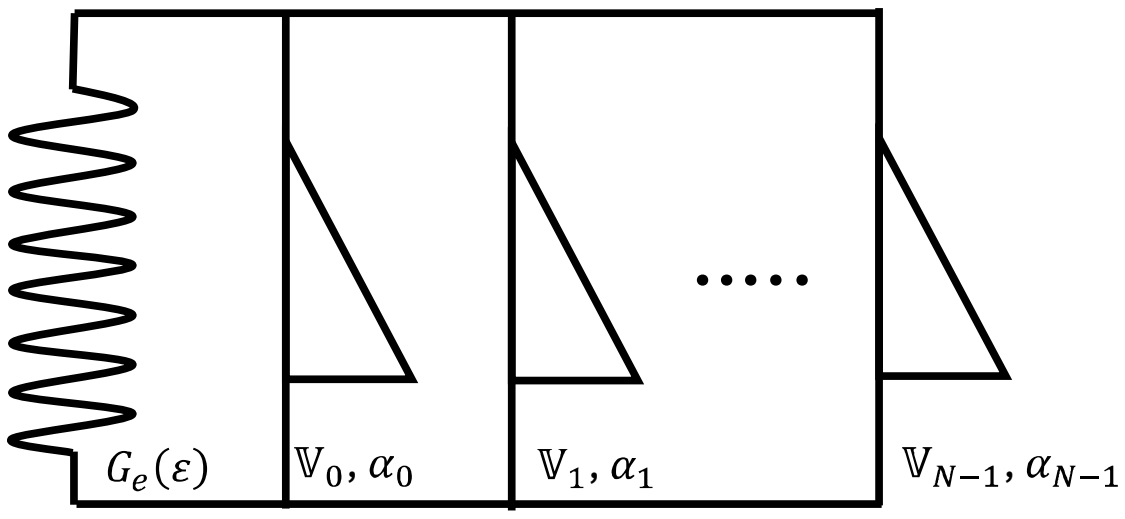

**Figure 2.** A parallel assembly of $N$ springpots and a spring, a fractional viscoelastic assembly proposed to model the post-gel states.

With this background, let us consider a general parallel fractional network assembly consisting of a single Hookean spring with modulus $G_e(\varepsilon)$ in parallel with $N$ springpots as shown in Figure 2. We label the springpots by the index $k$, where $k = 0$ to $N - 1$. Each $k$th springpot is characterized by a fractional order $\alpha_k \in (0, 1)$ and $\varepsilon$-dependent quasi-property $\mathbb{V}_k(\varepsilon)$. For the parallel assembly, the total strain across all elements is identical, which is equal to the applied strain $\gamma(t)$, and the total stress is the sum of stresses induced in each element. Consequently, under a step-strain input $\gamma(t) = \gamma_0 H(t)$, the constitutive relation of each springpot results in a relaxation modulus contribution $\mathbb{V}_k(\varepsilon)\, t^{-\alpha_k}/\Gamma(1 - \alpha_k)$. The Hookean spring, on the other hand, contributes the time-independent term $G_e(\varepsilon)$. The total relaxation modulus of the proposed network is therefore:

$$G(t,\varepsilon) = G_e(\varepsilon) + G_r(t,\varepsilon) = G_e(\varepsilon) + \sum_{k=0}^{N-1} \frac{\mathbb{V}_k(\varepsilon)}{\Gamma(1-\alpha_k)}\, t^{-\alpha_k}, \tag{82}$$

where $G_r(t,\varepsilon)$ is the relaxable part of the relaxation modulus. Eq. (82) has $2N+1$ unknowns: $N$ fractional orders $\{\alpha_k\}$ as well as quasi-properties $\{\mathbb{V}_k(\varepsilon)\}$, and one equilibrium modulus $G_e(\varepsilon)$. All of these parameters needed to be determined by the physical and scaling constraints mentioned in Table 1. Since the obvious constraints can be verified easily, we shall concentrate on those constraints that tangibly alter the model parameters.

As $\varepsilon \to 0$, the relaxation modulus given by Eq. (82) must recover the Winter–Chambon power-law expression: $\lim_{\varepsilon\to 0} G(t,\varepsilon) = S\, t^{-n}$, which leads to

$$G(t,0) = G_e(0) + \sum_{k=0}^{N-1} \frac{\mathbb{V}_k(0)}{\Gamma(1-\alpha_k)}\, t^{-\alpha_k} = S\, t^{-n} \tag{83}$$

For this equality to hold, we must have: 1. $G_e(0) = 0$, and 2. exactly one springpot, say $k = 0$, must have its exponent equal to $\alpha_0 = n$. All the other exponents $\alpha_k$ should have their values different from $n$ with corresponding $\mathbb{V}_k(\varepsilon)$, for all $k \geq 1$, tending to zero at $\varepsilon = 0$, or simply not contributing. Overall, these constraints lead to: $\alpha_0 = n$, $\mathbb{V}_0(0) = S\,\Gamma(1-n)$ and $\mathbb{V}_k(0) = 0$ for all $k \geq 1$. As $t \to \infty$, the relaxation modulus given by Eq. (82) approaches the equilibrium modulus $G_e(\varepsilon) = G_0\varepsilon^z$, validating the scaling law mentioned by Eq. (7) in Table 1.

We now apply a central constraint from Table 1 given by Eq. (13) that will lead to estimation of the fractional orders $\alpha_k$ and the quasi-properties $\mathbb{V}_k(\varepsilon)$. These constraints are: $G_r(t,\varepsilon) = St^{-n}\Psi\big(t/\tau_{max,G}(\varepsilon)\big)$, where $\Psi(0) = 1$ and $\Psi(\infty) = 0$ and $\left(\frac{\partial \ln\left(G_r(t,\varepsilon)\right)}{\partial \varepsilon}\right)_{\varepsilon\to 0} = B_G t^{\kappa_G}$. Since $G_r(t,0) = St^{-n}$, the combination of the above constraints intrinsically leads to $\left(\frac{\partial \Psi}{\partial \varepsilon}\right)_{\varepsilon\to 0} = B_G t^{\kappa_G}$. A Taylor expansion of the same about $\varepsilon = 0$ gives: $\Psi(t,\varepsilon) = 1 + B_G t^{\kappa_G}\varepsilon + O(\varepsilon^2)$. Therefore, the first-order nonlinear correction depends only on the combination, $B_G \varepsilon t^{\kappa_G}$. Taking this as a general expression and accounting for the constraints given by Eq. (13), the relaxation modulus can be written as:

$$G_r(t,\varepsilon) = S\,t^{-n}\sum_{k=0}^{\infty}\frac{\left(B_{G,k}\,\varepsilon\,t^{\kappa_G}\right)^k}{k!} = S\sum_{k=0}^{\infty}\frac{\left(B_{G,k}\,\varepsilon\right)^k}{k!}t^{k\kappa_G-n}, \tag{84}$$

where each mode $k$ carries its own independent coefficient $B_{G,k}$, such that $B_G$ here is the $k=1$ coefficient, i.e. $B_{G,1}=B_G$. This expression is structurally identical to the expression for the relaxation modulus given by Eq. (82), and the direct comparison leads to the fractional order being equal to $\alpha_k = n - k\kappa_G$, while the quasi-property becomes $\mathbb{V}_k(\varepsilon) = S(B_{G,k}\varepsilon)^k\Gamma(1-n+k\kappa_G)/k!$ and for it to remain non-negative for odd $k$, we must have $B_{G,k}\geq 0$. At this point, the expression given by Eq. (84) nominally contains infinitely many terms ($N\to\infty$). However, in order to have $G_r(t,\varepsilon)$ to be a non-increasing function of time, each springpot must have a fractional order strictly between zero and one: $0<\alpha_k<1$. For the above expression, the upper bound $\alpha_k<1$ is automatically satisfied since $\alpha_k = n-k\kappa_G \leq n < 1$ for all $k\geq 0$. However, the lower bound $\alpha_k>0$ is a necessary constraint leading to: $k<n/\kappa_G$. Consequently, the complete post-gel relaxation modulus of the parallel fractional network is given by:

$$G(t,\varepsilon) = G_0\varepsilon^z + S\sum_{k=0}^{\lfloor n/\kappa_G\rfloor}\frac{\left(B_{G,k}\varepsilon\right)^k}{k!}t^{k\kappa_G-n}, \tag{85}$$

where, throughout this paper, $\lfloor n/\kappa_G\rfloor$ denotes the largest integer strictly smaller than $n/\kappa_G$. It coincides with the usual floor function unless $n/\kappa_G$ is an integer, in which case $\lfloor n/\kappa_G\rfloor = n/\kappa_G - 1$, because $\alpha_k = n-k\kappa_G>0$ is strict (the same convention applies to $\lfloor n/\kappa\rfloor$ with $\kappa=\kappa_G$ in Post-Gel Model 2). Because the series is truncated at $k=K=\lfloor n/\kappa_G\rfloor$, $\Psi(x)$ grows as $x^{K\kappa_G}$ for $x\gg 1$, so the condition $\Psi(\infty)=0$ of Table 1 is not satisfied. However, $G_r$ still decays to zero, but only as $t^{-(n-K\kappa_G)}$. This is a limitation of the truncated post-gel models, and it applies equally to Post-Gel Model 2, which is algebraically identical to Model 1 mode by mode. Interestingly, the fractional network interpretation assigns to each term in the sum a distinct springpot with fractional order $\alpha_k = n-k\kappa_G$ and quasi-property $\mathbb{V}_k(\varepsilon) = S(B_{G,k}\varepsilon)^k\Gamma(1-n+k\kappa_G)/k!$, providing a mechanical framework for the post-gel relaxation.

In Eq. (85), unlike the exponential form and owing to the mode-specific coefficients $B_{G,k}$, there is no longer a single natural scaling variable $B_G\varepsilon\,t^{\kappa_G}$ that governs the entire relaxation dynamics. Rather, for each mode $k\geq 1$, there exists its own characteristic crossover timescale, which is identified by the condition that the $k$th term becomes of order unity:

$$\frac{\left(B_{G,k}\,\varepsilon\right)^k \tau_{max,G,k}^{k\kappa_G}}{k!} \sim 1 \Longrightarrow \tau_{max,G,k} = (k!)^{1/(k\kappa_G)}\left(B_{G,k}\,\varepsilon\right)^{-1/\kappa_G} = \tau_{G,k}\,\varepsilon^{-1/\kappa_G}, \tag{86}$$

where $\tau_{G,k} = (k!)^{1/(k\kappa_G)} B_{G,k}^{-1/\kappa_G}$ is the $\varepsilon$-independent material timescale associated with the mode $k$. Interestingly, despite the fact that the prefactor is mode-dependent, every mode diverges identically: $\tau_{max,G,k} \sim \varepsilon^{-1/\kappa_G}$ for all $k = 1,2,\dots,\lfloor n/\kappa_G \rfloor$, because all terms in the series possess the same time power $t^{\kappa_G}$. Furthermore, the dominant crossover timescale is the one that is closest to the critical gel state and is governed by the $k = 1$ mode. Since $B_{G,1} = B_G$ as per the constraint mentioned in Table 1, the primary longest relaxation time, which sets the crossover between power-law behavior associated with the critical gel and the long-time post-gel behavior, is therefore:

$$\tau_{max,G} = \tau_{max,G,1} = (B_G\,\varepsilon)^{-1/\kappa_G} = \tau_G\,\varepsilon^{-1/\kappa_G}, \tag{87}$$

where $\tau_G = B_G^{-1/\kappa_G}$ is the ε-independent timescale, similar to that in the exponential form. The higher modes k ≥ 2 set their own timescales $\tau_{max,\,G,\,k}$, which scale identically as $\varepsilon^{-1/\kappa_G}$ , differing only in their prefactors through $B_{G,k}$.

Next, we obtain the complex modulus by using the Fourier transform. Substituting the generalized series for $G_r$ given by Eq. (85) into Eq. (53) results in an expression for the complex modulus given by:

$$G^*(\omega,\varepsilon) = G_e(\varepsilon) + G_r^*(\omega,\varepsilon) = G_0\varepsilon^z + S\sum_{k=0}^{\lfloor n/\kappa_G \rfloor} \frac{\left(B_{G,k}\,\varepsilon\right)^k}{k!}\Gamma(1-n+k\kappa_G)(i\omega)^{n-k\kappa_G} \tag{88}$$

Separating real and imaginary parts using $(i\omega)^{n-k\kappa_G} = \omega^{n-k\kappa_G}\exp(i\pi(n-k\kappa_G)/2)$, the storage and loss moduli are:

$$G'(\omega,\varepsilon) = G_0\varepsilon^z + S\sum_{k=0}^{\lfloor n/\kappa_G \rfloor} \frac{\left(B_{G,k}\,\varepsilon\right)^k}{k!}\Gamma(1-n+k\kappa_G)\,\omega^{n-k\kappa_G}\cos\left(\frac{(n-k\kappa_G)\pi}{2}\right) \tag{89}$$

$$G''(\omega,\varepsilon) = S\sum_{k=0}^{\lfloor n/\kappa_G \rfloor} \frac{\left(B_{G,k}\,\varepsilon\right)^k}{k!}\Gamma(1-n+k\kappa_G)\,\omega^{n-k\kappa_G}\sin\left(\frac{(n-k\kappa_G)\pi}{2}\right) \tag{90}$$

In the above expressions, all frequency exponents $n - k\kappa_G$ are strictly positive for $k \leq \lfloor n/\kappa_G \rfloor$. Moreover, the successive terms contribute at progressively lower frequencies. In the limit $\varepsilon \to 0$, all terms with $k \geq 1$ vanish and both moduli reduce exactly to the Winter–Chambon critical gel expressions given by Eqs. (3) and (4), satisfying the continuity condition mentioned in Table 1. The major difference from the expression for the post-gel state proposed by Joshi [40] and Eqs. (89) and (90) is that the $\varepsilon$-dependence of each term at $k \geq 2$ is now governed by an independent coefficient $B_{G,k}$ that provides additional flexibility in fitting the full $\omega$ as well as $\varepsilon$ dependence of the experimental data away from the critical gel state.

Next, we differentiate $G_r'(\omega, \varepsilon)$ and $G''(\omega, \varepsilon)$ with respect to $\varepsilon$. Note that the $k = 0$ term is independent of $\varepsilon$ while for $k \geq 2$, the general term is proportional to $\varepsilon^k$, and its derivative tends to 0 as $\varepsilon \to 0$: $k\varepsilon^{k-1} \to 0$. Therefore, only the $k = 1$ term survives in the limit $\varepsilon \to 0$:

$$\left.\frac{\partial G_r'}{\partial \varepsilon}\right|_{\varepsilon \to 0} = S\, B_G\, \Gamma(1 - n + \kappa_G)\, \omega^{n-\kappa_G} \cos\left(\frac{(n-\kappa_G)\pi}{2}\right), \tag{91}$$

$$\left.\frac{\partial G''}{\partial \varepsilon}\right|_{\varepsilon \to 0} = S\, B_G\, \Gamma(1 - n + \kappa_G)\, \omega^{n-\kappa_G} \sin\left(\frac{(n-\kappa_G)\pi}{2}\right), \tag{92}$$

where $B_G = B_{G,1}$. This suggests that near the critical gel state the derivatives of the dynamic moduli are governed entirely by the $k = 1$ coefficient $B_{G,1} = B_G$. The higher-order coefficients $(k \geq 2)$ that are present in the generalized form manifest only at $O(\varepsilon^2)$ and beyond and hence influence the behavior of the material progressively farther from the critical gel state. Interestingly, the experimentally accessible near-critical-gel signatures remain unchanged. Dividing Eqs. (91) and (92) and by $G'$ and $G''$ at the critical gel state given by Eqs. (3) and (4) and leads to the logarithmic derivatives that scale as $\sim\omega^{-\kappa_G}$ validating Eq. (18) of Table 1. The corresponding $C_{G\to S} = \left.\frac{\partial \ln G_r'}{\partial \ln G''}\right|_{\varepsilon \to 0} = \cot\left(\frac{(n-\kappa_G)\pi}{2}\right)\tan\left(\frac{n\pi}{2}\right)$, derived (Eq. (A15)) in Appendix A and summarized in Eq. (24) of Table 1, is independent of any constitutive model.

**Post-Gel Model 2. Multimodal Prabhakar Model**

In the pre-gel analysis, the two-parameter Mittag-Leffler function, i.e. the dashpot-free limit of Pre-Gel Model 1, was extended to the three-parameter Mittag-Leffler (Prabhakar) function, which we designated Pre-Gel Model 2. In the same spirit, we propose the following expression for the post-gel relaxation modulus in terms of Prabhakar function given by:

$$G(t,\varepsilon) = G_0\varepsilon^z + S\Gamma(1-n)t^{-n}\sum_{j=1}^{P} w_j\ E_{\kappa_G,1-n}^{\gamma_G}\left(\vartheta_{G,j}\varepsilon\, t^{\kappa_G}\right), \tag{93}$$

where $w_j \geq 0$, $\sum_{j=1}^{P} w_j = 1$, $\vartheta_{G,j} > 0$ are mode-specific post-gel material constants, $\gamma_G > 0$ is the Prabhakar exponent, and $\overline{\vartheta_G^k} \equiv \sum_{j=1}^{P} w_j\, \vartheta_{G,j}^k$ denotes the $k$th weighted moment. Note that the pre-gel Prabhakar model has a negative argument $E_{\kappa_S,1-n}^{\gamma_S}\left(-\vartheta_{S,j}\varepsilon t^{\kappa_S}\right)$, owing to which $G(t,\varepsilon)$ is completely monotonic and $G(t,\varepsilon) \to 0$ as $t \to \infty$. Furthermore, it also leads to: $\left(\frac{\partial \ln G(t,\varepsilon)}{\partial \varepsilon}\right)_{\varepsilon\to 0} = -B_S t^{\kappa_S}$. For the post-gel states, contrary to the pre-gel states, an increase in $\varepsilon$ increases the degree of crosslinking. Therefore, for the post-gel model, it is necessary to use a positive argument $+\vartheta_{G,j}\varepsilon t^{\kappa_G}$, which leads to $(\partial \ln G_r/\, \partial\varepsilon)_{\varepsilon\to 0} = B_G t^{\kappa_G}$ as necessitated by the constraint mentioned in Table 1. With a positive argument $(+\vartheta_{G,j}\varepsilon t^{\kappa_G})$, however, the Prabhakar function grows without bound as $t \to \infty$, which is similar to that observed for the exponential function as discussed by Joshi and coworkers [11, 40]. Consequently, the full expression of the Prabhakar function must be expanded in a series and appropriately truncated to keep the relaxation modulus a non-increasing function of time. In principle, Eq.(93) can also be expressed in the Havriliak–Negami form. However, since it eventually needs truncating, we represent it as just the truncated Prabhakar series.

Expressing the Prabhakar function in Eq.(93) as an expansion, the proposed relaxation modulus is given by:

$$G_r(t,\varepsilon) = S\, t^{-n}\sum_{k=0}^{\infty}\frac{\Gamma(\gamma_G+k)\Gamma(1-n)}{k!\,\Gamma(\gamma_G)\Gamma(\kappa_G k+1-n)}\overline{\vartheta_G^k}\varepsilon^k\, t^{k\kappa_G}. \tag{94}$$

For this expression, it can be easily shown that for $k \geq n/\kappa$ the term $\sim t^{(k\kappa-n)}$ does not decay: it is constant for $k = n/\kappa$ and increases for $k > n/\kappa$. Consequently, to ensure that $G_r(t,\varepsilon)$ is a non-increasing function of $t$ that vanishes at long times (the equilibrium modulus being carried by $G_0\varepsilon^z$ alone), we must limit the series given in Eq. (94) from $k = 0$ to $\lfloor n/\kappa \rfloor$, and the resulting expression for the complete post-gel relaxation modulus for the multimodal Prabhakar model is given by:

$$G(t,\varepsilon) = G_0\varepsilon^z + S\, t^{-n}\sum_{k=0}^{\lfloor n/\kappa_G\rfloor}\frac{\Gamma(\gamma_G+k)\Gamma(1-n)}{k!\,\Gamma(\gamma_G)\Gamma(\kappa_G k+1-n)}\overline{\vartheta_G^k}\varepsilon^k\, t^{k\kappa_G}. \tag{95}$$

As in Post-Gel Model 1, the truncation at $k = K = \lfloor n/\kappa \rfloor$ means that $\Psi(\infty) = 0$ is not satisfied, although $G_r$ still decays as $t^{-(n-K\kappa)}$. This is a limitation of the truncated post-gel models.

Having established the above expression for the relaxation modulus in the form of a truncated Prabhakar function, it is instructive to examine its relationship with Post-Gel Model 1 before performing any further analysis. We can compare the two expressions for $G(t,\varepsilon)$, Eq. (85) of Post-Gel Model 1 and Eq. (95) of Post-Gel Model 2, term by term. A careful analysis of both the expressions suggests that Post-Gel Model 2 is algebraically identical to Post-Gel Model 1 under the following mode-by-mode substitution:

$$B_{G,k} = \left[\frac{\Gamma(\gamma_G + k)\Gamma(1 - \mathrm{n})}{\Gamma(\gamma_G)\Gamma(\kappa_G k + 1 - n)}\right]^{1/k} \tag{96}$$

for $k = 0$ to $\lfloor n/\kappa_G \rfloor$, where $\overline{\vartheta_G^k} = \sum_{j=1}^{P} w_j \vartheta_{G,j}^k$. This identification clearly suggests that the Post-Gel Model 2 is a constrained subset of Post-Gel Model 1. Consequently, every solution of Post-Gel Model 2 can be expressed as a solution of Post-Gel Model 1 via the substitution given by Eq. (96). However, the converse does not hold in general: the parameters $B_{G,k}$ generated by Post-Gel Model 2 are not mutually independent but are coupled through the Prabhakar exponent $\gamma_G$ as well as the modes $\{w_j, \vartheta_{G,j}\}$.

It should be noted that, in the case of the pre-gel state, as shown by Eq. (65), $\gamma_S$ governs the evolving polydispersity of the total relaxable finite clusters. The post-gel exponent $\gamma_G$, in Eq. (95), on the other hand, provides a necessary physical degree of freedom for the relaxable fraction of the percolated system. It is important to note that the relaxable part of the relaxation modulus constitutes only the finite modes and clusters, and not the infinite network. Therefore, $\gamma_G$ can be considered as a phenomenological measure of the structural heterogeneity and topological hindrance of these trapped finite clusters as well as dangling ends. By modulating the modal weights of the surviving relaxation modes, $\gamma_G$ physically captures how this relaxable fraction delays the macroscopic transition to the permanent elastic plateau. Furthermore, the second term of Eq. (95) is limited to $k = 0$ to $\lfloor n/\kappa \rfloor$ to preserve monotonicity of the relaxation process. As $n$ decreases, the network becomes denser at the critical gel state. Physically this makes the relaxation of the dangling clusters more sluggish. Mathematically, this leads to $\lfloor n/\kappa \rfloor = 1$. These aspects are expected to have a significant effect on the value of $\gamma_G$.

Furthermore, a direct consequence of the fact that post-gel Model 2 is a constrained subset of post-gel Model 1 is that the complex modulus, storage and loss moduli, their derivatives and logarithmic derivatives with respect to $\varepsilon$, and the proportionality constant $C_{G\to S}$, follow the corresponding results of post-gel Model 1, with the substitution $B_{G,k}$ given by Eq. (96) . Moreover, all functional forms, frequency scaling, and near-critical-gel signatures are identical between the two models. The scientific merit of post-gel Model 2 relative to post-gel Model 1, therefore, is not in producing new scaling behavior but in replacing the mutually independent amplitudes $\{B_{G,k}\}$ of the latter with physically structured parameters $\{w_j, \vartheta_{G,j}\}$ and the Prabhakar exponent $\gamma_G$.

## Continuity at the Critical Gel Point

As mentioned by Joshi [40], in order to have a physically consistent description of the sol-gel transition, the linear viscoelastic properties must vary continuously as the material passes through the critical gel point. The continuity must hold not only for the values of the properties but also for their first derivatives with respect to $p$. As stated in Table 1, this requirement of the continuity of derivatives constitutes the most compelling physical constraint connecting the pre-gel and post-gel state descriptions. Since $\varepsilon = p_c - p$ in the pre-gel regime and $\varepsilon = p - p_c$ in the post-gel regime, the derivative with respect to $p$ transforms according to $\frac{\partial G^*}{\partial p} = -\frac{\partial G^*}{\partial \varepsilon}$ for $p < p_c$ in the pre-gel states, while $\frac{\partial G^*}{\partial p} = +\frac{\partial G^*}{\partial \varepsilon}$ for $p > p_c$ in the post-gel states.

From the results derived for the pre-gel models, the derivatives of the dynamic moduli with respect to $p$ in the limit $p \to p_c^-$ are given by Eqs. (59) and (60) for pre-gel Model 1 and Eqs. (80) and (81) for pre-gel Model 2 with $\partial/\,\partial p = -\,\partial/\,\partial\varepsilon$. Note that both the sets of derivatives for the pre-gel state are identical. The corresponding derivatives for the post-gel models in the limit $p \to p_c^+$ are given by Eqs. (91) and (92) of post-gel Model 1 with $\partial/\,\partial p = \partial/\,\partial\varepsilon$. The continuity of the first derivatives of the dynamic moduli at $p_c$ requires the left-hand and right-hand limits of the derivatives to be identical and given by:

$$\left(\frac{\partial G'}{\partial p}\right)_{p\to p_c^-} = \left(\frac{\partial G'}{\partial p}\right)_{p\to p_c^+} \text{ and } \left(\frac{\partial G''}{\partial p}\right)_{p\to p_c^-} = \left(\frac{\partial G''}{\partial p}\right)_{p\to p_c^+} \tag{97}$$

Since the frequency dependences on the two sides of the above-mentioned expressions for the derivatives are $\omega^{n-\kappa_S}$ and $\omega^{n-\kappa_G}$ respectively, and since power-law functions of distinct exponents are linearly independent of each other for all $\omega > 0$, the equality mentioned by Eq. (97) can hold for all frequencies simultaneously if and only if the frequency exponents match, resulting in:

$$\kappa_S = \kappa_G = \kappa \tag{98}$$

This is the symmetry condition mentioned in Table 1 by Eq. (22). This result suggests that symmetric evolution of the relaxation dynamics on both sides of the critical gel state, which was also derived by Joshi [40] through conventional viscoelastic framework, is a universal, model-agnostic consequence of the requirement that the derivatives of the dynamic moduli are continuous across the critical gel point. This symmetry is therefore a property of the intrinsic scaling structure, which is entirely independent of the modal architecture or the specific form of the fractional function employed. The molecular origin of this symmetry can be traced to the common percolation geometry that governs the relaxation dynamics on both sides of the gelation transition.

In the pre-gel state, the longest relaxation time is set by the largest finite cluster, whose size is the percolation correlation length, $\xi(\varepsilon) \sim \varepsilon^{-\nu}$, where $\nu$ is the correlation-length exponent [5, 75, 76]. In colloidal gels with short-range attraction, a nonequilibrium percolation process gives $\nu \approx 0.8$, consistent with percolation theory [6], although the particle attachment and detachment steps do not obey detailed balance. With a dynamic exponent $D$, $\tau_{\max,S} \sim \xi^{D} \sim \varepsilon^{-\nu D}$, which gives $1/\kappa_S = \nu D$. On the other side of the gelation transition, in the post-gel state, the longest relaxing units are the unattached clusters trapped within the infinite network backbone. Note that the gelation dynamics that produces the largest pre-gel as well as post-gel clusters represents two complementary facets of the same critical percolation geometry, which are infinitesimally separated on the scale of $p$ on either side of the critical value $p_c$. The correlation length diverges with the same exponent $\nu$ on both sides, so with the same $D$ we have $\tau_{\max,G} \sim \xi^{D} \sim \varepsilon^{-\nu D}$ and $1/\kappa_G = \nu D = 1/\kappa_S$. A single correlation length governs the size of the dominant relaxing entities on either side of the gel point, whether these are pre-gel or post-gel finite clusters. Therefore, the symmetry of the relaxation dynamics on both sides of the gelation transition: $\kappa_S = \kappa_G = \kappa$ is not an imposed mathematical condition but a microstructural necessity. This argument holds for any $\nu$ and $D$. So does the relation $n = z/(z+s)$ obtained below from continuity [40], which contains neither $\nu$ nor the dimension $d$. However, the values of $\nu$ and $D$ follow only from additional closures. Each of those rests on its own assumption, so they are alternative model representations and need not hold together. (i) If the modulus is $k_B T$ per correlation volume, $G_e \sim k_B T\, \xi^{-d}$ (the blob form for networks [48], applied here to the percolation correlation length as a scaling assumption), then $z = d\nu$, $D = d/n$ and, with $n = z/(z+s)$, $n = d\nu/(d\nu + s)$ [9]. (ii) In the electrical analogy of de Gennes, $z$ equals the conductivity exponent [76], $z = (d-2)\nu + \zeta_R$ with $\zeta_R \approx 1$ is the resistance exponent of percolation theory; for $d = 3$ this gives $\nu \approx z - 1$ and $D \approx 1/(n - \kappa)$. (iii) Relations between $n$ and the fractal dimension $d_f$, such as those of Muthukumar [41] and Zaccone et al. [42] discussed in the Introduction, rest on specific cluster models. For $\zeta_R = 1$, closures (i) and (ii) coincide only at $\nu = 1/2$; for $d = 3$ and $\nu = 0.88$ they give $z = 2.64$ and $1.88$. None of them is used in the models or fits of this work.

With Eq. (22) or Eq. (98) reestablished, the coefficient matching further leads to:

$$B_S = B_G = B_{G,1}, \tag{99}$$

confirming that the pre-gel and post-gel material constants given by respectively Eq. (14) and Eq. (15), which also govern the leading-order departure from the critical state are identical. This is the fractional viscoelasticity counterpart of the condition established by Joshi [40] for the conventional viscoelastic framework. As emphasized by Joshi [40] and mentioned in the Introduction, this symmetry condition has a profound implication for the scaling relations. From the pre-gel analysis we have $\kappa_S = (1-n)/s$ while from the post-gel analysis we have $\kappa_G = n/z$; consequently, the equality $\kappa_S = \kappa_G$ yields: $n = z/(z+s)$, which is the hyperscaling relation tabulated in Table 1 Eq. (23). Therefore, as in the conventional framework, the fractional viscoelastic analysis establishes that any material undergoing a gelation transition must satisfy the symmetry condition and hyperscaling relations. Finally, the symmetry condition obtained in Appendix A also leads to $C_{S\to G} = C_{G\to S} = C = \cot\left(\frac{(n-\kappa)\pi}{2}\right)\tan\left(\frac{n\pi}{2}\right)$ resulting in $\left(\frac{\partial G''}{\partial G'}\right)_{\varepsilon\to 0} = \tan\left(\frac{(n-\kappa)\pi}{2}\right)$. The parameter $C$ therefore appears as a single, unified frequency-independent material fingerprint irrespective of the direction of approach to the critical gel state. In addition, this work also emphasizes that parameter $C$ is model-agnostic in nature. This further confirms that $C$ is an intrinsic characteristic of the critical gel state governed solely by $n$ and $\kappa$.

**Results and Discussion**

In the previous section we developed the theoretical framework through fractional viscoelasticity and the multi-parameter Mittag-Leffler-Prabhakar function. In this framework, we derive a generalized set of model expressions for both the time-domain relaxation modulus $G(t,\varepsilon)$ as well as the frequency-domain dynamic moduli ($G'(\omega,\varepsilon)$ and $G''(\omega,\varepsilon)$) across the pre-gel and post-gel states. In this section, to establish the physical validity of these models, we rigorously evaluate these models against canonical experimental datasets documented in the literature. A central aspect of this work is the strict symmetry of the underlying relaxation dynamics as the material passes through the critical gel state, in either direction. Consequently, throughout the remainder of this paper, as per Eq. (98), we use a single value of the relaxation scaling exponent $\kappa$ for both the pre-gel and post-gel regimes. In order to assess the proposed framework across distinct topological classes of gels, we begin by examining time-domain relaxation modulus data associated with the sol-gel transition for a chemically crosslinked poly(dimethylsiloxane) (PDMS) system that shows a critical relaxation exponent of $n = 0.45$. Subsequently, we fit the model to the frequency-domain dynamic moduli data for a physically associating poly(vinyl alcohol) (PVA) system characterized by $n = 0.77$ over a broad frequency range. Since these two systems possess very different critical network characteristics, in addition to their widely separated $n$ values, they collectively provide a benchmark experimental dataset for validating the proposed phenomenological models.

It should be noted that the exponents $n$, $\kappa$, $s$ and $z$ are not predicted by the present models. Theory relates them to the correlation-length exponent $\nu$ and the dimension $d$ only through model-dependent closures (see "Continuity at the Critical Gel Point"), which are not used here. They are, however, related through the hyperscaling relations: $s = (1-n)/\kappa$ and $z = n/\kappa$ (Table 1), so only two of them are independent. We obtain $n$ and $\kappa$ by fitting the models to experimental data. The exponent $n$ is obtained from the linear viscoelastic response at the critical gel state, while $\kappa$ is obtained from the approach to it. Exponent $z$ is obtained from the post-gel fit and can be compared with $n/\kappa$. The exponents $s$ and hence the Prabhakar exponent $\gamma_S = s$, then follow, and can be tested against independently measured $\eta_0(\varepsilon)$ and $G_e(\varepsilon)$. Pre-Gel Model 1 is admissible only for $n + \kappa < 1$, whereas Pre-Gel Model 2 and the post-gel models admit any pair $0 < \kappa < n < 1$. All other parameters ($S$, $B_S$, $\vartheta_{S,j}$, $w_j$, $G_0$, $\gamma_G$, $b$) are obtained by fitting model to the data or are fixed by the continuity (symmetry) constraint.

Obtaining high-quality, time-domain relaxation modulus data in the vicinity of the critical gel point is challenging, and such datasets remain scarce in the literature. In this work we use the PDMS dataset obtained by Winter and coworkers [78, 79]. This data was generated by meticulously arresting the chemical crosslinking reaction of PDMS at various stages of network formation (quantified by the degree of crosslinking). Winter and coworkers [78, 79] performed linear viscoelastic experiments on these samples by adjusting the applied strain amplitude in the range 3 to 0.005 as the network stiffened gradually. The dynamic moduli across an extensive temperature window from −50°C to 180°C were measured in the linear range and, using the time–temperature superposition principle, they constructed robust master curves at a reference temperature of 34℃, spanning five decades of reduced frequency. Finally, they determined discrete relaxation spectra from these master curves [78, 79] and calculated from them the time-domain relaxation modulus $G(t, \varepsilon)$ that spans the pre-gel, critical-gel, and post-gel states. We plot this experimental data in Fig. 3. The plot depicts the critical gel state ($\varepsilon = 0$) represented by the dash-dotted line with a constant slope of $-n$ on a logarithmic scale. The data on the left and right side of the critical gel state are those associated with the pre-gel and post-gel states, respectively.

We employ the following procedure to fit the experimental data. We first consider the pre-gel states described by Pre-Gel Model 1, the exact response of the series assembly of two springpots and a dashpot (Eq. (50)), shown as dashed lines in Fig. 3, together with the post-gel states described by the truncated Prabhakar model (Post-Gel Model 2, Eq. (95)). The fitted values of $\varepsilon$ are listed in the legend of Fig. 3 and the other parameters in Table 2. Since $\lfloor n/\kappa \rfloor = 1$, only one term in the series of Eq. (85) and Eq. (95) survives, so Post-Gel Model 1 (parallel arrangement of springpots and a spring) and Post-Gel Model 2 (truncated Prabhakar function) become mathematically identical. Pre-Gel Model 1 has only one parameter beyond the exponents $n = 0.45$ and $\kappa = 0.29$, the quasi-property gel strength $S$ and the constant $B_S = 76.0\ \mathrm{s}^{-\kappa}$, namely the dimensionless dashpot strength $b = 0.19$ (equivalently $\mu_0 = 353\ \mathrm{s}^{n-1}$ or $\eta_S = 1.60$ Pa·s). Eq. (14) then fixes $\vartheta_\varepsilon = B_S \Gamma(\kappa + 1 - n)/\Gamma(1 - n) = 52.8\ \mathrm{s}^{-\kappa}$ and $\tau_S = \vartheta_\varepsilon^{-1/\kappa} = 1.15 \times 10^{-6}$ s. With $b = 0.19$, the dashpot dominates beyond $t_* \approx b^{-1/(1-n-\kappa)} \tau_{\max,S} \approx 6 \times 10^2\ \tau_{\max,S}$, so Model 1 departs from its dashpot-free limit only at long times for the larger values of $\varepsilon$, which is where $b$ is constrained. For the smallest $\varepsilon$ the terminal regime lies beyond the experimental window. The pre-gel data are also fitted with the multimodal Prabhakar formulation (Pre-Gel Model 2, Eq. (65)), shown as solid lines in Fig. 3, with the same post-gel fit. Its two modes ($P = 2$), with weights $w_j$ and timescale factors $\vartheta_{S,j}$, are listed in Table 3. $B_S$, $n$, $\kappa$ and $S$ for this fitting are the same as those for Model 1 fitting. As shown in Fig. 3, both combinations of pre-gel and post-gel models agree well with the experimental data. Note that the fits in Fig. 3 are not independent fits of the respective pre-gel and post-gel models but strictly adhere to the continuity of derivative of the dynamic moduli (Eq. (97)) at the critical gel state given by constraints Eqs. (98) and (99), which underscores their profound physical validity.

Interestingly, the fitted values of $\varepsilon$ differ between the two pre-gel models, since $\varepsilon$ is a fitted parameter that enters each model through a model-specific product: $\vartheta_\varepsilon \varepsilon$ for Model 1 and $\vartheta_{S,j} \varepsilon$ for Model 2. The two rates are linked through the first-order constraint, Eq. (69), which gives $\vartheta_\varepsilon = \gamma_S \overline{\vartheta_S^1}$, where $\overline{\vartheta_S^1} = \sum_j w_j \vartheta_{S,j} = 27.8\ \mathrm{s}^{-\kappa}$ (Table 3). Both fits use the same $B_S$ and hence the same $\varepsilon$ scale. Placing the crossover at the same time requires $\varepsilon_1 = \varepsilon_2 \overline{\vartheta_S^1}/\vartheta_\varepsilon = \varepsilon_2/\gamma_S \approx 0.53\ \varepsilon_2$, consistent with the fitted ratios 0.51, 0.51 and 0.63. Requiring equal zero-shear viscosities, $\eta_S^{(1)} \varepsilon_1^{-s} = \eta_S^{(2)} \varepsilon_2^{-s}$, where $\eta_S^{(1)}$ and $\eta_S^{(2)}$ denote the viscosity prefactor η_S of Pre-Gel Model 1 (the dashpot constant) and of Pre-Gel Model 2 (Eq. (72) and the definition following it), respectively, with $\mathcal{A}_{n,\kappa_S} = \sum_j w_j \vartheta_{S,j}^{-s}$, gives $\varepsilon_1/\varepsilon_2 = 0.54$. Both matching conditions apply only when the crossover and the terminal regime lie within the experimental window. As $\varepsilon \to 0$, these regimes move out of the window, and the two models, which share the same first-order response as shown by Eq. (69), are expected to require the same $\varepsilon$, so that $\varepsilon_1/\varepsilon_2 \to 1$.

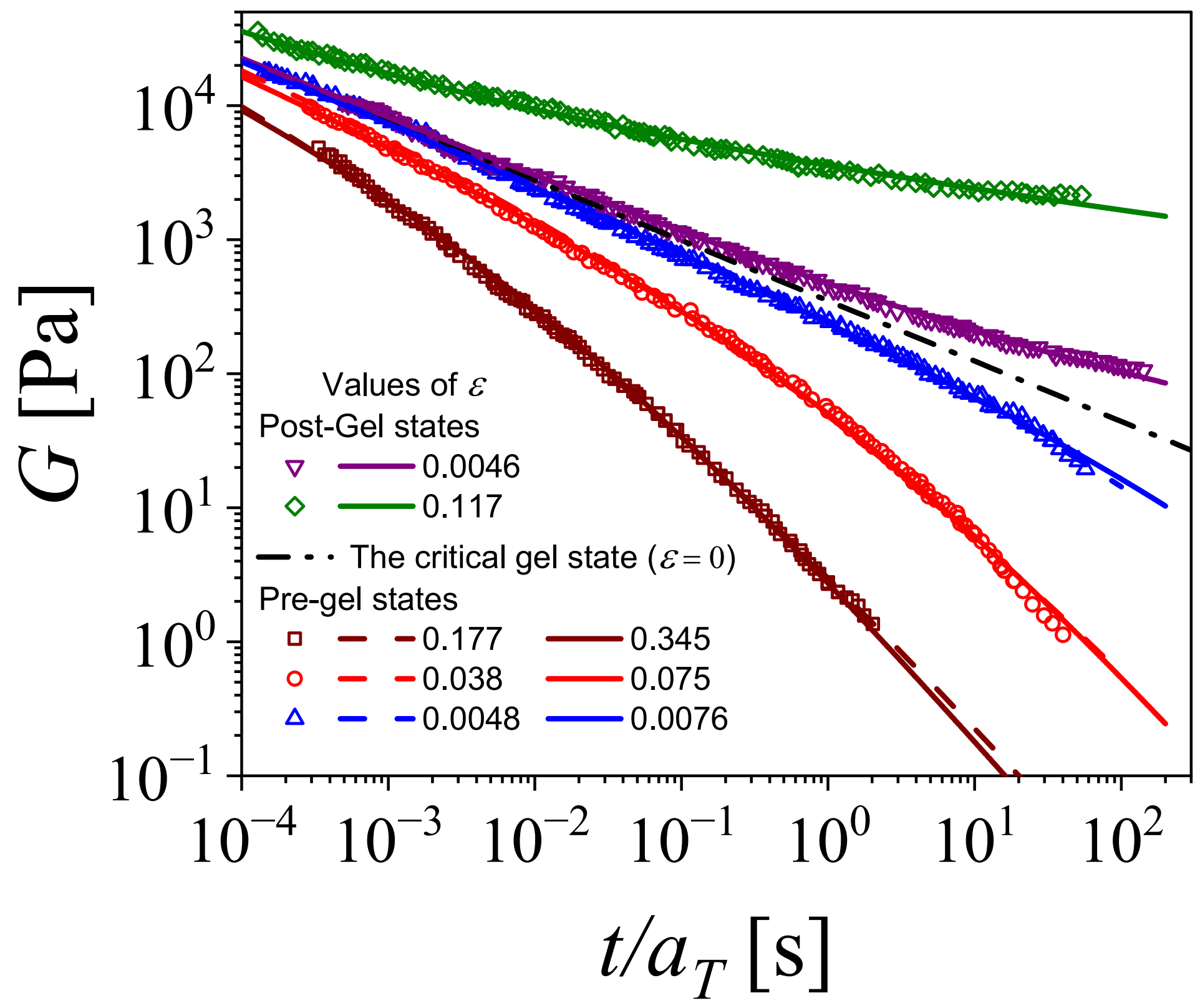

**Figure 3.** Relaxation modulus $G$ [Pa] for a chemically crosslinked PDMS system is plotted as a function of reduced time $t/a_T$ [s] at varying distances $\varepsilon$ from the critical gel point. The experimental data are taken from Winter and coworkers [78, 79]. The symbols represent the transformed experimental data extracted from the time-temperature superposition for the pre-gel as well as the post-gel states. The dash-dotted line describes the pure scale-free power-law relaxation characteristic of the critical gel state ($\varepsilon = 0,\ n = 0.45$). Post-gel states: solid lines, Post-Gel Model 2 (truncated multimodal Prabhakar expression), which is identical to Post-Gel Model 1 here because $\lfloor n/\kappa \rfloor = 1$; this fit is common to both pre-gel models. Pre-gel states: solid lines, Pre-Gel Model 2 (multimodal Prabhakar expression); dashed lines, Pre-Gel Model 1 (exact response of two springpots and a dashpot in series, Eq. (50)). The fitted values of $\varepsilon$ are listed in the legend and differ between the two pre-gel models, because $\varepsilon$ enters each model through a model-specific rate (see text). The relaxation dynamics are constrained symmetrically across the transition through $\kappa = 0.29$. Note that the pre- and post-gel expressions are solved simultaneously, linked through the continuity constraints Eqs. (98) and (99). All model parameters are listed in Table 2 and 3.

**Table 2. Parameters of Pre-Gel Model 1 and Post-Gel Model 2 for the PDMS system**

| Parameter | Value |
|---|---|
| Critical exponent ($n$) | 0.45 |
| Relaxation scaling exponent ($\kappa$) | 0.29 |
| Critical gel stiffness ($S$) | 350.166 $\mathrm{Pas}^n$ |
| Symmetry constraint ($B_S = B_{G,1}$) | 76.0 $\mathrm{s}^{-\kappa}$ |
| Dimensionless dashpot strength ($b$) | 0.19 |
| Dashpot constant ($\mu_0$) | 353 $\mathrm{s}^{n-1}$ |
| Equilibrium modulus scale ($G_0$) | 3677.22 Pa |
| Equilibrium modulus exponent ($z$) | 1.55 |
| $\gamma_G \vartheta_G^1$ | 52.8 $\mathrm{s}^{-\kappa}$ |
| $\varepsilon$, pre-gel | 0.0048, 0.038, 0.177 |
| $\varepsilon$, post-gel | 0.0046, 0.117 |

**Table 3. Parameters of Pre-Gel Model 2 for the PDMS system‡**

| Parameter | Value |
|---|---|
| $\gamma_S = s$ | 1.897 |
| Mode 1 weight ($w_1$) | 0.767 |
| Mode 1 timescale factor ($\vartheta_{S,1}$) | 10.455 $s^{-\kappa}$ |
| Mode 2 weight ($w_2$) | 0.233 |
| Mode 2 timescale factor ($\vartheta_{S,2}$) | 85.064 $s^{-\kappa}$ |
| $\varepsilon$, pre-gel | 0.0076, 0.075, 0.345 |

‡ The post-gel parameters are common to both pre-gel models and are therefore listed in Table 2. $\gamma_S = s$ belongs to Model 2 only.

For PDMS, $\lfloor n/\kappa \rfloor = 1$, so only the product $\gamma_G \vartheta_G^1$ enters the post-gel model, through $B_{G,1} = \gamma_G \vartheta_G^1 \Gamma(1-n)/\Gamma(1-n+\kappa)$ (Eq. (96) with k = 1). The parameters $\gamma_G$ and $\vartheta_G$ therefore cannot be determined separately from these data. The value 52.8 in Table 2 corresponds to the choice $\vartheta_G = 1s^{-\kappa}$ and has no independent physical content; only $\gamma_G \vartheta_G^1 = B_S \Gamma(1-n+\kappa)/\Gamma(1-n) = 52.8\, s^{-\kappa}$ is determined. A physical interpretation of $\gamma_G$ requires data with $K \geq 2$, where the higher-order terms lift this degeneracy. In Fig. 4, we plot storage modulus $G'(\omega)$ and loss modulus $G''(\omega)$ for an aqueous poly(vinyl alcohol) (PVA) solution as functions of angular frequency $\omega$ for different distances from the critical gel state in the pre-gel as well as the post-gel regime. The experimental data shown in Fig. 4 are taken from Joshi et al. [22] on PVA solution undergoing a thermoreversible gel–sol transition. Unlike the chemically crosslinked PDMS system analyzed in Fig. 3, where irreversible crosslinks formed through the covalent bonds lead to a permanent sol–gel transition, PVA solution system forms physically associating gels in which the network junctions are thermoreversible. Specifically, the crosslink zones in PVA solution system that get created due to a decrease in temperature are composed of microcrystalline domains supported by hydrogen bonding between the hydroxyl groups of adjacent PVA chains. These network junction zones nucleate and grow upon cooling and span the space. Upon further cooling, the network becomes progressively dense in the post gel regime. Upon heating the network junction zones dissolve reversibly, restoring the sol state at sufficiently high temperatures. Consequently, the PVA solution system traverses the critical gel state reversibly in either direction of temperature change, making it a convenient model system for probing the universal features of the gelation transition [5].

On this PVA solution system, Joshi et al. [22] performed oscillatory frequency sweep experiments spanning $\omega$ =0.1–100 rad/s at a shear stress amplitude of 1 Pa, which lies within the linear viscoelastic regime at all temperatures investigated. The authors identified the critical gel temperature to be $T_C = 20$°C, at which both $G'$ and $G''$ exhibit the characteristic Winter–Chambon power-law scaling with identical frequency exponents as shown in Fig. 4(e). The experimental data shown in Fig. 4 correspond to the gel–sol transition obtained by heating the PVA solution through $T_C = 20$°C. The post-gel data for four lower temperatures (12 to 18°C) are shown in Figs. 4(a–d), while the pre-gel data for four higher temperatures (22 to 28°C) are plotted in Figs. 4(f–i). We quantify the distance from the critical point by $\varepsilon = |T - T_C|/T_C$. with $T$ and $T_C$ in °C ($T_C = 20$ °C). Another temperature scale only rescales $\varepsilon$ by a constant, which is absorbed in $B_S$, $\vartheta_S$, $j$, $B_{G,k}$ and $G_0$ (see below).

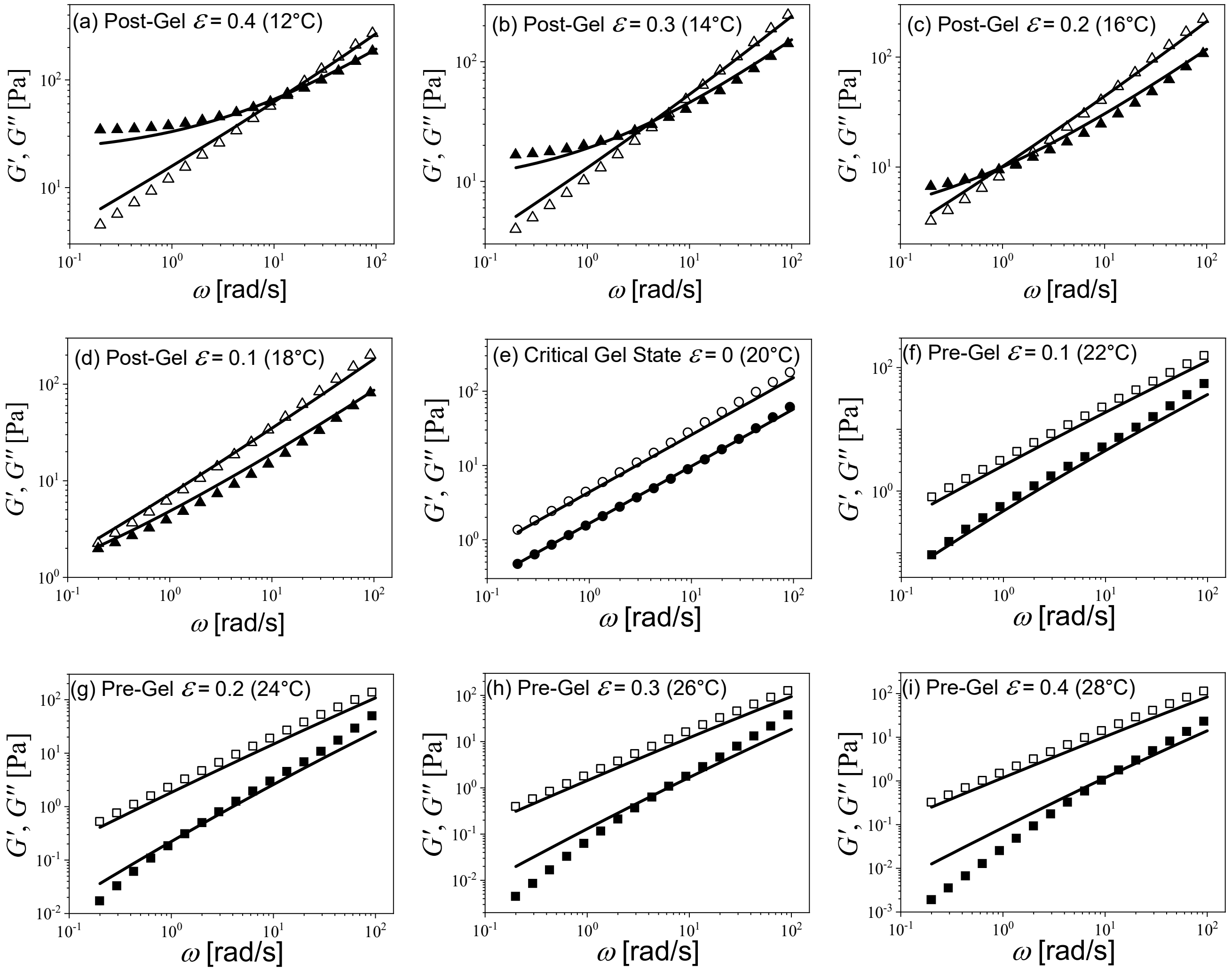

**Figure 4.** Storage modulus $G'(\omega)$ and loss modulus $G''(\omega)$ for an aqueous poly(vinyl alcohol) (PVA) solution, as it undergoes gel-sol transition, are plotted as functions of angular frequency $\omega$, at various distances ($\varepsilon$) from the critical gel point: in the post-gel states (a) 0.4, (b) 0.3, (c) 0.2, (d) 0.1, the critical gel state (e) 0 and the pre-gel states (f) 0.1, (g) 0.2, (h) 0.3 and (i) 0.4. Experimental data (symbols) are taken from Joshi et al. [22], obtained during the gel–sol transition by heating through the critical temperature $T_c$ =20°C. In all the panels, the experimental data of $G'$ and $G''$ are fitted using Pre-Gel Model 2 (multimodal Prabhakar formulation, Eqs. (78) and (79)) and Post-Gel Model 1 (parallel fractional network, Eqs. (89) and (90)). The critical gel state experimental data is fitted by either of the models in a limit of $\varepsilon \to 0$. Note that the pre- and post-gel state expression for the dynamic moduli are solved simultaneously, coupled through the continuity constraint Eqs. (98) and (99). All model parameters are listed in Table 4.

In all panels of Fig. 4, the experimental $G'$and $G''$data are fitted using Pre-Gel Model 2 (the multimodal Prabhakar formulation) described by Eqs. (78) and (79), along with the Post-Gel Model 1 (the parallel fractional network model) represented by Eqs. (89) and (90). At $\varepsilon = 0$, both the pre-gel and post-gel formulations reduce consistently to the Winter–Chambon power-law scaling expressions described by Eqs. (3) and (4). It is important to emphasize that the expressions for the dynamic moduli in the pre-gel and post-gel regimes are not fitted independently to the experimental data but are solved simultaneously while being coupled through the continuity constraints given by Eqs. (98) and (99). The complete set of model parameters that lead to the fits shown in Fig 4 is provided in Table 4. It can be seen that the models provide an excellent fit to the experimental data in the immediate vicinity of the critical gel state, corresponding to relatively small values of $\varepsilon$. However, as $\varepsilon$ increases in either direction, pre-or post-gel, away from the critical state, the quality of the fits gradually declines. This deterioration is more pronounced for the pre-gel fits than for their post-gel counterparts. Such behavior is physically expected, since the proposed expressions are most accurate in the neighborhood of the critical gel state. Furthermore, Fig. 4 also shows that in general the model fits $G''$ data better than $G'$ data. The ability of the model to quantitatively reproduce the experimental response naturally weakens as the system evolves farther away from criticality, deep into the sol or the dense gel state. In Fig. 4, we do not employ the Pre-Gel Model 1 (the two-springpots-in-series with a dashpot) owing to the admissibility constraint associated with it. As mentioned above, Pre-Gel Model 1 is valid only for systems satisfying: $n + \kappa < 1$. For the PVA solution system, we have, $n = 0.77$ and $\kappa = 0.2655$, resulting in $n + \kappa = 1.0355 > 1$, which violates the admissibility condition. On the other hand, for the post-gel data we do not use Post-Gel Model 2, because it does not fit well. In Post-Gel Model 1 the fit returns $B_{G,2} \approx 0$ (Table 4). That is, the data require only the $k = 0$ and $k = 1$ terms. In Model 2, the $k = 2$ amplitude is tied to $\gamma_G$ and $\vartheta_G$ through Eq. (96) and cannot be made negligible while $B_{G,1}$ is held fixed.

**Table 4. Model Parameters for the PVA Solution System**

| **Universal Parameters** | **Value** |
|---|---|
| Critical exponent ($n$) | 0.77 |
| Relaxation scaling exponent ($\kappa$) | 0.26552 |
| Critical gel stiffness ($S$) | 1.184 Pa s$^n$ |
| Equilibrium modulus scale ($G_0$) | 287.9 Pa |
| Equilibrium modulus exponent ($z$) | 2.9 |

| Other Parameters | | |
|---|---|---|
| Symmetry constraint ($B_S = B_{G,1}$) | | 19.01 s$^{-\kappa}$ |
| Pre-Gel Modes **($P = 1$)** | Mode 1 Weight ($w_1$) | 1 |
| | Mode 1 Timescale Factor ($\vartheta_{S,1}$) | 9.909 s$^{-\kappa}$ |
| Post-Gel, $B_{G,2}$ | | $1.5 \times 10^{-7}$ s$^{-\kappa}$ |

Interestingly, in the models developed in this paper, for a given state, the distance from the critical gel state $\varepsilon$, is not an independent parameter with a unique value. Evidently, $\varepsilon$ always appears in the expression of relaxation modulus and dynamic moduli through the specific invariant products. In the post-gel regime, the $k$th term of the finite series depends exclusively on the product $\left(B_{G,k}\, \varepsilon\right)^k$. In addition, the equilibrium modulus term in the post-gel expression depends on the composite term $G_0 \varepsilon^z$. In the pre-gel Prabhakar series expression, the analogous invariant is $\vartheta_{S,1}\varepsilon$, which appears as $\left(\vartheta_{S,1}\, \varepsilon\right)^k$ in the $k$th term. Furthermore, the continuity constraint $B_S = B_{G,1}$ operates entirely at the level of these products. Since both sides possess the same $\varepsilon$-dependence, the continuity constraint $\left(B_S = B_{G,1}\right)$ is preserved under any rescaling. This indicates that $\varepsilon$ is free to take any positive value say from $\varepsilon_{\text{old}}$ to $\varepsilon_{\text{new}}$ provided that $B_{G,1}$, $B_{G,2}$, $B_S$, and $\vartheta_{S,1}$ are simultaneously multiplied by $\varepsilon_{\text{old}}/\varepsilon_{\text{new}}$ so as to keep their respective invariant products constant. In addition, $G_0$ needs to be multiplied by $(\varepsilon_{\text{old}}/\varepsilon_{\text{new}})^z$ to preserve $G_0\varepsilon^z$. Note that the parameters $S, n, \kappa, z$, and $\gamma_S$ are structurally decoupled from $\varepsilon$, and hence remain strictly unchanged. In short, the above-mentioned products involving $\varepsilon$ form closely coupled pairs that uniquely determine the state of the system either in the pre- or post-gel domain.

Collectively, the experimental validations presented in Figs. 3 and 4 establish the physical consistency and broad applicability of the fractional viscoelastic framework as well as the Prabhakar function-based approach for two distinct gel-forming material classes: chemically crosslinked PDMS undergoing sol-gel transition and thermoreversible PVA solution undergoing gel-sol transition. The proposed frameworks are also applicable across two complementary experimental modes: the time-domain relaxation modulus and the frequency-domain dynamic moduli at different distances from the critical gel state. Importantly, in both systems, the pre-gel and post-gel model expressions are solved simultaneously, strictly coupled through the continuity of the first derivatives of the dynamic moduli at the critical gel state. Imposition of this condition confirms the symmetry condition $\kappa_S = \kappa_G = \kappa$ and the hyperscaling relation $n = z/(z + s)$ as theoretical necessities rather than coincidental observations.

A complementary question worth addressing is origin of power-laws that are followed by various properties as a transition point is approached, with the jamming transition as an analogous microscopic example. Zaccone et al. [80, 81] recently proposed a theoretical framework on jamming transition that starts from a particle–bath Hamiltonian and expresses the shear modulus as the affine (Born) modulus minus a nonaffine correction that is fixed by the vibrational density of states and the affine-force correlator. For athermal soft spheres it gives a shear modulus that vanishes linearly as the mean coordination number $Z$ approaches the isostatic value $Z_c = 2d$ from above, leading to $G \propto Z - 2d$, where $d$ is the dimensionality of space [80, 81]. The same Hamiltonian gives a viscosity below jamming that diverges because the lower edge of the vibrational spectrum closes as $2d - Z$, with $\eta \propto (2d - Z)^{-1}$ to $(2d - Z)^{-1.3}$ depending on the numerical exponent adopted for the spectrum edge [81]. Interestingly this pair of power-laws on the two sides of the transition is reminiscent of the gelation transition. However, the mechanisms in both cases are different. Jamming is athermal and set by isostaticity, whereas gelation is set by connectivity percolation and a diverging cluster size. The comparison also underlines the scope of the present work. Fractional viscoelasticity, as used here, is a mathematically convenient representation of scale-free spectra and models how various properties diverge in a power-law fashion as the critical gel state is approached. Microscopic theories, on the other hand, analyze why such spectra and power-law exponents emerge from the underlying many-body dynamics. Consequently, a gel analog of the nonaffine calculation, for a percolating network with central-force and bond-bending interactions, would be needed to analyze what affects the values of $n$, $\kappa$, $z$ and $s$, which in the present work are taken from experimental data.

## Conclusions

In this work, we develop a family of fractional viscoelastic and Prabhakar-function-based models for materials undergoing sol-gel transitions. The prior literature on application of fractional viscoelasticity to the gelation transition applied it either at the critical gel point in isolation or independently on each side of the transition. The present framework resolves this by incorporating the continuity of complex modulus and its first derivatives with respect to the degree of crosslinking as an explicit constraint. For the pre-gel regime, two models are proposed.

Pre-Gel Model 1, a series assembly of two springpots and a dashpot, is solved exactly: its complex modulus is given in closed form, its relaxation modulus as a bivariate Mittag-Leffler series, and its viscosity diverges with the correct exponent $s$. Its first-order dependence on $\varepsilon$, which enters all constraints of Table 1, is carried by the second springpot alone. It is restricted to systems satisfying $n+\kappa<1$. Its dashpot-free limit is the two-parameter Mittag-Leffler function, which has no finite viscosity. Pre-Gel Model 2, on the other hand, employs the three-parameter Mittag-Leffler–Prabhakar function, which lifts the restriction $n+\kappa<1$ and has a finite viscosity built in, while preserving all required scaling properties. A key result of Pre-Gel Model 2 is the unique determination of the Prabhakar exponent, $\gamma_S=\frac{1-n}{\kappa}=s$, which identifies $\gamma_S$ with the viscosity divergence exponent and makes it experimentally measurable. This is a consistency result: it restates the Winter–Chambon exponent through $n=1-\kappa\gamma_S$ and the scaling relation Eq. (10). Mechanical assemblies are therefore not necessary for the pre-gel description, since the Prabhakar form satisfies every constraint without one, and the simplest pre-gel assembly is not sufficient, since it excludes gels with $n+\kappa\geq 1$. Springpots remain exact at the critical gel, and the post-gel parallel assembly is exact for the truncated series. Remarkably, Prabhakar-function-based pre-gel relaxation modulus and complex modulus correspond exactly to the classical Havriliak–Negami relaxation function of dielectric spectroscopy. Interestingly, $\kappa_S$ and $\gamma_S$ take the roles of the two shape parameters in the Havriliak–Negami function. Furthermore, unlike in dielectric analysis, where these are empirical fitting parameters, in the present analysis they are not just experimentally measurable but acquire explicit physical meaning. Here the Havriliak–Negami function, which is ubiquitous in dielectric relaxation, arises as an exact expression for the pre-gel complex modulus whose two shape parameters are fixed by the gelation exponents.

For the post-gel regime, we propose Post-Gel Model 1 as a parallel assembly of a Hookean spring and $N$ springpots. The Hookean spring leads to a finite equilibrium modulus while the springpots produce a power-law relaxation spectrum. Post-Gel Model 2, on the other hand, is based on the Prabhakar function. The series representation of both models needs truncation to ensure that the relaxation modulus is a non-increasing function of time. Interestingly, Post-Gel Model 2 is algebraically equivalent to Post-Gel Model 1 through a mode-by-mode identification of the amplitude coefficients. The two post-gel models, therefore, share identical scaling behavior, with Post-Gel Model 1 providing a greater parametric flexibility for experimental data validation.

The main constraint imposed across the two sets of models, in the pre-gel and post-gel states, is the continuity of $G'$ and $G''$ and their first derivatives with respect to the degree of crosslinking at the critical gel point. Enforcement of this continuity constraint leads to the following three consequences that are model-agnostic in nature: (i) the symmetry condition: $\kappa_S=\kappa_G=\kappa$, suggesting identical scaling of the relaxation dynamics on both sides of the transition; (ii) the equality of the leading-order departure coefficients in the expressions for the pre- and post-gel regimes; and (iii) the hyperscaling relation $n = z/(z+s)$, independent of $\nu$ and $d$, which appears as a theoretical necessity imposed by the requirement of the continuous evolution of viscoelastic properties rather than as an empirical coincidence.

We validate the proposed models against experimental data from two canonical systems with widely differing network topology and critical exponents. For chemically crosslinked PDMS undergoing sol-gel transition $(n = 0.45, \kappa = 0.29)$, the model quantitatively reproduces the time-domain relaxation modulus across multiple pre-gel and post-gel states. For thermoreversible PVA solution undergoing gel-sol transition $(n = 0.77, \kappa = 0.2655)$, on the other hand, the model describes well the frequency-domain dynamic moduli across a broad range of temperatures on either side of the critical gel point. In the case of the PVA solution system, the fit quality is observed to deteriorate at large $\varepsilon$, which is consistent with the asymptotic character of the model. It is important to note that, in both cases, the pre-gel and post-gel fits are performed simultaneously, coupled through the continuity constraints. A final result of greater significance is that the expression: $(\partial G''/\partial G')_{\varepsilon\to 0} = \tan((n-\kappa)\pi/2)$ is model-agnostic and frequency-independent. We believe that this expression constitutes an intrinsic rheological fingerprint of the critical gel state determined exclusively by $n$ and $\kappa$. Intriguingly, its structural analogy with the loss tangent at the critical gel state: $tan\delta = tan(n\pi/2)$, through the substitution $n \to n-\kappa$, reveals a profound internal symmetry of the gelation transition. The present framework thus offers a unified and experimentally testable description of the evolution of linear viscoelastic properties across the gelation transition.

## Appendix A. Model-agnostic determination of $\left(\frac{\partial G''(\omega,\varepsilon)}{\partial G'(\omega,\varepsilon)}\right)_{\varepsilon\to 0}$ through the Kramers–Kronig relations

In Table 1, we mention the proportionality constant $C = \left(\frac{\partial \ln G'(\omega,\varepsilon)}{\partial \ln G''(\omega,\varepsilon)}\right)_{\varepsilon\to 0}$ and the symmetry condition $\kappa_S = \kappa_G = \kappa$, which can be derived for any viscoelastic model that adheres to the critical gelation theory. In this appendix, we show that the expression for $C$ follows directly from the Kramers–Kronig relations, without invoking any particular constitutive model.

### A1. Power-law viscoelasticity and Kramers–Kronig relations

To Begin with, let us consider a standard linear viscoelastic response in terms of complex modulus: $G^*(\omega) = G'(\omega) + iG''(\omega)$, for which the Kramers–Kronig relation is given by [82]:

$$G'(\omega,\varepsilon) - G_e(\varepsilon) = -\frac{2\omega^2}{\pi}\int_0^{\infty} \frac{G''(\xi,\varepsilon)}{\xi(\xi^2-\omega^2)} d\xi \tag{A1}$$

$$G''(\omega,\varepsilon) = \frac{2}{\pi}\int_0^{\infty} \omega \frac{G'(\xi,\varepsilon) - G_e(\varepsilon)}{\xi^2-\omega^2} d\xi \tag{A2}$$

Now let us assume a power-law functional form for $G^*(\omega)$ for $\omega > 0$, given by:

$$G^*(\omega) = G_e + A\,\omega^m e^{i\varphi}, \tag{A3}$$

where $A > 0$ is real, $m$ is real and $\varphi$ is an unknown constant. Eq. (A3) can be resolved into real and imaginary parts leading to:

$$G'(\omega) = G_e + A\omega^m \cos\varphi, \tag{A4}$$

$$G''(\omega) = A\omega^m \sin\varphi. \tag{A5}$$

Incorporation of Eqs. (A4) and (A5) independently into Eqs. (A1) and (A2) results in the same answer, given by:

$$\varphi = \frac{\pi m}{2} \tag{A6}$$

Therefore, the complex modulus satisfying the relationship $G^*(\omega) = G_e + A\,\omega^m e^{i\varphi}$ can be written as:

$$G^*(\omega) - G_e = A(i\omega)^m = A\omega^m e^{\frac{i\pi m}{2}}. \tag{A7}$$

Eqs. (A3) and (A7), taken together, are a very profound result, which implies that when the complex modulus shows a power-law dependence on frequency, the phase angle between the real and imaginary parts is necessarily fixed by the power-law exponent. For the Winter-Chambon criterion: $G'(\omega) \sim \omega^n$ and $G''(\omega) \sim \omega^n$, which inevitably leads to $\tan\delta = \tan\left(\frac{n\pi}{2}\right)$ at the critical gel state.

**A2. Linear viscoelasticity in the vicinity of the gelation transition**

Now let us consider the evolution of linear viscoelasticity in the vicinity of the gelation transition. The frequency-domain scaling of the logarithmic derivative of the complex modulus in the pre-gel states (Table 1, Eq. (16)) is,

$$\left(\frac{\partial \ln(G^*(\omega,\varepsilon))}{\partial \varepsilon}\right)_{\varepsilon\to 0} = (\beta_S' + i\beta_S'')\omega^{-\kappa_S} = A_S\omega^{-\kappa_S}e^{i\varphi_S}, \tag{A8}$$

while that in the post-gel states (Table 1, Eq. (17)) is:

$$\left(\frac{\partial \ln(G^*(\omega,\varepsilon) - G_e(\varepsilon))}{\partial \varepsilon}\right)_{\varepsilon\to 0} = (\beta_G' + i\beta_G'')\omega^{-\kappa_G} = A_G\omega^{-\kappa_G}e^{i\varphi_G}. \tag{A9}$$

We multiply Eqs. (A8) and (A9) by the critical-gel complex modulus, $G^*(\omega,\varepsilon\to 0) = S\Gamma(1-n)(i\omega)^n$ given by Eq. (2). Since $i^n = e^{i\frac{n\pi}{2}}$, for the pre-gel states we have:

$$\left(\frac{\partial G^*(\omega,\varepsilon)}{\partial \varepsilon}\right)_{\varepsilon\to 0} = A_S S\Gamma(1-n)\omega^{n-\kappa_S}e^{i(\frac{n\pi}{2}+\varphi_S)}, \tag{A10}$$

while for the post-gel state it gives:

$$\left(\frac{\partial (G^*(\omega,\varepsilon) - G_e(\varepsilon))}{\partial \varepsilon}\right)_{\varepsilon\to 0} = A_G S\Gamma(1-n)\omega^{n-\kappa_G}e^{i(\frac{n\pi}{2}+\varphi_G)}. \tag{A11}$$

Differentiation of Eqs. (A1) and (A2) with respect to $\varepsilon$ in the limit $\varepsilon \to 0$ results in:

$$\left(\frac{\partial (G'(\omega,\varepsilon) - G_e(\varepsilon))}{\partial \varepsilon}\right)_{\varepsilon\to 0} = -\frac{2\omega^2}{\pi}\int_0^\infty \frac{1}{\xi(\xi^2-\omega^2)}\left(\frac{\partial G''(\xi,\varepsilon)}{\partial \varepsilon}\right)_{\varepsilon\to 0} d\xi, \tag{A12}$$

$$\left(\frac{\partial G''(\omega,\varepsilon)}{\partial \varepsilon}\right)_{\varepsilon\to 0} = \frac{2}{\pi}\int_0^\infty \frac{\omega}{\xi^2-\omega^2}\left(\frac{\partial (G'(\xi,\varepsilon) - G_e(\varepsilon))}{\partial \varepsilon}\right)_{\varepsilon\to 0} d\xi. \tag{A13}$$

Just as Eqs. (A1), (A2) and (A3) lead to Eq. (A6), Eqs. (A10), (A12) and (A13) result in $n\pi/2 + \varphi_S = (n-\kappa_S)\pi/2$, which can be simplified to:

$$\varphi_S = -\frac{\kappa_S\pi}{2}. \tag{A14}$$

Substituting Eq. (A14) into Eqs. (A8) and (A10), separating the resulting expressions into their real and imaginary components, and subsequently taking their ratio results in:

$$C_{S\to G} = \left(\frac{\partial \ln G'(\omega,\varepsilon)}{\partial \ln G''(\omega,\varepsilon)}\right)_{\varepsilon\to 0} = \cot\left(\frac{(n-\kappa_S)\pi}{2}\right)\tan\left(\frac{n\pi}{2}\right) \tag{A15}$$

$$\left(\frac{\partial G''(\omega,\varepsilon)}{\partial G'(\omega,\varepsilon)}\right)_{\varepsilon\to 0} = \tan\left(\frac{(n-\kappa_S)\pi}{2}\right) \tag{A16}$$

Similarly, for the post-gel states, Eqs. (A11)–(A13) yield:

$$\varphi_G = -\frac{\kappa_G\pi}{2}. \tag{A17}$$

Following the same procedure as for the pre-gel states, substituting Eq. (A17) into Eqs. (A9) and (A11), resolving the resulting expressions into their real and imaginary parts, and taking their ratio gives:

$$C_{G\to S} = \left(\frac{\partial \ln G'(\omega,\varepsilon)}{\partial \ln G''(\omega,\varepsilon)}\right)_{\varepsilon\to 0} = \cot\left(\frac{(n-\kappa_G)\pi}{2}\right)\tan\left(\frac{n\pi}{2}\right) \tag{A18}$$

$$\left(\frac{\partial G''(\omega,\varepsilon)}{\partial G'(\omega,\varepsilon)}\right)_{\varepsilon\to 0} = \tan\left(\frac{(n-\kappa_G)\pi}{2}\right) \tag{A19}$$

Since the changes in the storage and loss moduli on either side of the critical gel state must be continuous as the system undergoes the gelation transition, we have $C_{S\to G} = C_{G\to S} = C$. Comparing Eqs. (A15) and (A18), and canceling the common factor $\tan(n\pi/2)$, this condition reduces to:

$$\cot\left(\frac{(n-\kappa_S)\pi}{2}\right) = \cot\left(\frac{(n-\kappa_G)\pi}{2}\right). \tag{A20}$$

Since the cotangent is monotonic and one-to-one over the physically admissible range of its argument, Eq. (A20) can hold only if

$$\kappa_S = \kappa_G = \kappa. \tag{A21}$$

This is precisely the symmetry condition of Table 1. In this appendix, we obtain it from the Kramers–Kronig relations applied directly to the $\varepsilon$-derivatives of the complex modulus, independently of any constitutive model. Since $\kappa_S$ and $\kappa_G$ govern, respectively, the pre-gel and post-gel divergence of the longest relaxation time as mentioned in Table 1, their equality validates the symmetry condition. Through it, the hyperscaling relation $n = z/(z+s)$ is also verified as a consequence of Kramers–Kronig relations and continuity of the response, rather than of any particular model.

In Fig. A1, we plot $(\partial G''/\partial G')_{\varepsilon\to 0}$ given by Eqs. (A16) or (A19) as a function of the critical relaxation exponent $n$ for various representative values of the relaxation scaling exponent $\kappa$. Several important physical features of $(\partial G''/\partial G')_{\varepsilon\to 0}$ are apparent from Fig. A1. First, for any fixed $\kappa$, $(\partial G''/\partial G')_{\varepsilon\to 0}$ is a monotonically increasing function of $n$. As $n$ increases, the system possesses progressively more liquid-like power-law character at the critical gel state. Fig. A1 suggests that in such a limit, the loss modulus becomes increasingly sensitive relative to the storage modulus as the material transitions through the critical gel point, and $(\partial G''/\partial G')_{\varepsilon\to 0}$ grows as $n \to 1$. On the other hand, for fixed $n$, $(\partial G''/\partial G')_{\varepsilon\to 0}$ decreases monotonically with increasing $\kappa$. Note that $\kappa$ is known to have a value in the vicinity of 0.2. In addition, as per Eqs. (8) and (9): $\tau_{\max,S} = \tau_S\, \varepsilon^{-1/\kappa}$ and $\tau_{\max,G} = \tau_G\, \varepsilon^{-1/\kappa}$ suggest that an increase in κ results in a less intense divergence of the relaxation dynamics on both sides of the critical gel state, which could be responsible for the suppression of the differential evolution of $G''$ relative to $G'$. Furthermore, $(\partial G''/\partial G')_{\varepsilon\to 0}$ is strictly positive throughout the entire physically admissible parameter space: $\kappa < n < 1$. This is a direct consequence of the physical constraint $n > \kappa$ specified by Joshi [40], which originates from the requirement that the evolution of equilibrium modulus should remain finite at the critical gel state. The consequence of $(\partial G''/\partial G')_{\varepsilon\to 0}$ remaining always positive is that if $G'$ increases (decreases) as a material traverses the critical gel state, $G''$ also must increase (decrease) in the same direction. Furthermore, the very fact that $(\partial G'/\partial p)_{p\to p_c} > 0$ suggests that both $G'$ and $G''$ must necessarily increase with the degree of crosslinking as the sol transforms into a gel and vice versa.

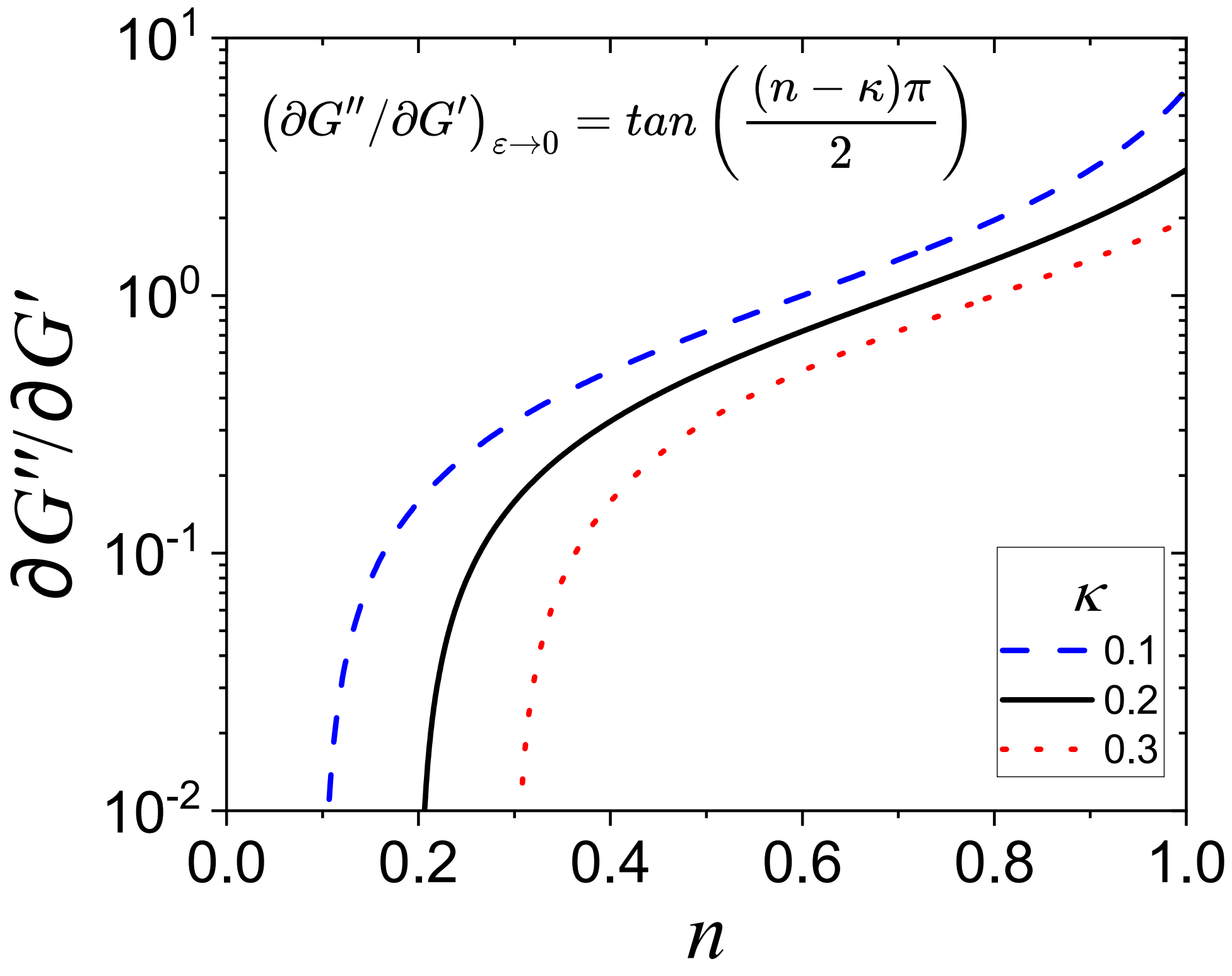

**Figure A1.** Relative change in $G''$ with respect to $G'$ as the system passes through the critical gel state, $(\partial G''/\partial G')_{\varepsilon\to 0}$, is plotted as a function of the critical relaxation exponent $n$ for representative values of the relaxation scaling exponent $\kappa$. It is important to note that this expression is model-agnostic and is independent of frequency. Note that the region $n \leq \kappa$ is physically inaccessible as discussed by Joshi [40].

It is worth commenting on the resemblance between two quantities obtained through completely independent routes: ratio of $G''$ to $G'$ at the critical gel state, $\tan\delta = (G''/G')_{\varepsilon=0} = \tan(n\pi/2)$, and the derivative of $G''$ with respect to $G'$ at the critical gel state: $(\partial G''/\partial G')_{\varepsilon\to 0} = \tan[(n-\kappa)\pi/2]$. Interestingly, while both are frequency-independent, they probe fundamentally different aspects of the transition. The former $(\tan\delta)$ is a purely structural fingerprint that depends only on $n$, and describes the fractal, self-similar architecture of the weakest percolated network exactly at the critical gel point. On the other hand, $(\partial G''/\partial G')_{\varepsilon\to 0}$ is a dynamical fingerprint suggesting how $G''$ evolves with respect to $G'$ as the system undergoes gelation transition. The astonishing aspect is that the dynamical response $((\partial G''/\partial G')_{\varepsilon\to 0})$ is expressed in the same tangent-of-half-angle form as the static structural one $((G''/G')_{\varepsilon=0})$, merely shifting the angle by an amount set by how fast the system evolves. This shift is what distinguishes a fixed structural state of the critical gel from a dynamic approach toward or away from it.

**Acknowledgment:** The author acknowledges financial support from the Anusandhan National Research Foundation, Government of India (Grant number: JCB/2022/000040). The author thanks Prof. Gareth McKinley and Mohua Das for fruitful discussions.